# Understanding Human-COVID-19 Dynamics using Geospatial Big Data: A Systematic Literature Review

Binbin Lin[a], Lei Zou [a*], Mingzheng Yang[a], Bing Zhou[a], Debayan Mandal[a], Joynal Abedin[a], Heng Cai[a], Ning Ning[b]

[a]*Department of Geography, Texas A&M University, College Station, TX, USA;*

[b]*Department of Statistics, Texas A&M University, College Station, TX, USA*

**Email: lzou@tamu.edu*

The COVID-19 pandemic has dramatically changed human daily life. To mitigate the pandemic's impacts, different countries and regions implemented various policies to contain COVID-19 and residents showed diverse responses. These human responses in turn shaped the uneven spatial-temporal spread of COVID-19. Consequently, the human-pandemic interaction is complex, dynamic, and interconnected. Delineating the reciprocal effects between human society and the pandemic is imperative for mitigating risks from current and future epidemics. Geospatial big data acquired through mobile applications and sensor networks have facilitated near-real-time tracking and assessment of human responses to the pandemic, enabling a surge in researching human-pandemic interactions. However, these investigations involve inconsistent data sources, human activity indicators, relationship detection models, and analysis methods, leading to a fragmented understanding of human-pandemic dynamics. To assess the current state of human-pandemic interactions research, we conducted a synthesis study based on 67 selected publications between March 25th, 2020, and January 9th, 2023. We extracted key information from each article across six categories, e.g., publication details, research area and time, data, methodological framework, and results and conclusions. Results reveal that regression models were predominant in relationship detection, featured in 67.16% papers. Only two papers employed spatial-temporal models, notably underrepresented in the existing literature. Studies examining the effects of policies and human mobility on pandemic's health impacts were the most prevalent, each comprising 12 articles (17.91%). Only 3 papers (4.48%) delved into bidirectional interactions between human responses and the COVID-19 spread. These findings shed light on the need for future research to spatially and temporally model the long-term, bidirectional causal relationships within human-pandemic systems to inform evidence-based, hyperlocal pandemic mitigation strategies.

## 1. Introduction

The COVID-19 pandemic has dramatically changed human daily life and affected societies in much of the world. To tackle the uneven impacts of the pandemic, different countries and regions implemented distinct policies to contain the pandemic (e.g., lockdown, business closure, and distance education). Residents showed diverse responses and compliance with COVID-19 policies (e.g., wearing masks, maintaining social distancing, etc.) (Jun et al., 2021). The discrepancies in responding policies and behaviors in turn shaped the disparate patterns of COVID-19 spread across space and time (Bryant & Elofsson, 2020). Therefore, human-pandemic interaction is a complex, dynamic, interconnected system with feedback across social, economic, environmental, health, and pandemic dimensions. There is a need to delineate the reciprocal effects between human society and the pandemic. This information will help formulate strategies to mitigate the risk of future epidemics.

Geospatial big data collected from remote sensing, social media, cell phone apps, vehicles, and sensor networks enable near-real time, multi-dimensional tracking and evaluating human perceptions, sentiments, and behaviors toward the pandemic (Effenberger et al., 2020). Incorporating human-centric information derived from geospatial big data into spatial epidemiology can facilitate the visualization, analysis, and prediction of the impacts of the pandemic on human societies and the ensuing influences of human dynamics on epidemics.

Nevertheless, interpreting human behaviors from geospatial big data is challenging. First, geospatial big data contain a sheer amount of noise information irrelevant to human behaviors during the pandemic. Second, geospatial big data collected from diverse sources have varying formats, structures, and standards, making it difficult to integrate them for analysis. Third, different research adopted various indicators based on geospatial big data to measure human activities, leading to inconsistent analysis methods and sometimes contradictory results (Razzaq et al., 2022).  This discrepancy hinders reliable knowledge discovery and comparative studies. Fourth, only a small proportion of geospatial big data are embedded with precise locations, leaving the majority of the data unusable for spatial analysis directly. Ultimately, geospatial big data cannot represent the entire population (Lin et al., 2024), leading to biases and inaccurate estimations of human activities.

Uncovering the dynamics of the human-pandemic system poses several additional challenges. First, the human-pandemic system is a complex adaptive system with multiple interacting factors such as COVID-19 policies, human behaviors (e.g., mobility, public awareness and sentiment toward COVID-19), socioeconomic and demographic characteristics, healthcare availability and accessibility, and environmental conditions (Hsiang et al., 2020). It is difficult to consider all these factors in epidemiological modeling. Second, the relationships in this complex system vary across epidemic stages and geographic regions (Cinarka et al., 2021; Wallin Aagesen et al., 2022), and contain numerous feedback loops. Quantitatively describing, analyzing, and interpreting these complex casual relationships necessitates advanced spatial-explicit models empowered by machine learning or deep learning. Third, two prominent issues exist in spatially and temporally analyzing human-pandemic dynamics, the Modifiable Areal Unit Problem (MAUP) (Wong, 2004) and the Modifiable Temporal Unit Problem (MTUP) (Cheng & Adepeju, 2014). These issues underscore the sensitivity of analysis outcomes to variations in spatial-temporal scales hierarchy and zonal systems. Finally, the uncertainties surrounding the pandemic such as vaccinations and new variants can further complicate the human-pandemic relationships.

This study conducted a synthesis study on research that uses geospatial big data to understand human-COVID-19 interactions. Published literature between the March 25th, 2020 and January 9th, 2023, were included. Two key questions underpin this research. First, how did existing efforts leverage geospatial big data in human-COVID-19 dynamics research? Second, how did COVID-19 and human responses interact across different pandemic phases and space? To address these questions, we outlined four objectives: (1) to summarize the common geospatial datasets and indicators for monitoring the COVID-19 spread and human responses; (2) to identify traditional and advanced models for detecting relationships in human-pandemic dynamics; (3) to elucidate the evolving interactions and geographical disparities of human-pandemic dynamics; and (4) to highlight gaps in the current frameworks and knowledge of human-pandemic dynamic systems and propose future research directions that can advance the field and improve pandemic management.

## 2. Methods of Literature Retrieval and Analysis

The conceptual framework of this study is illustrated in Figure 1, which underscores the reciprocal impact between COVID-19 and human responses. Initially, the spread of COVID-19 dramatically affected human behaviors, including the implementation of diverse COVID-19 policies, uneven public awareness and sentiment toward COVID-19, and changes in human mobility. These human responses interacted with each other, and in turn, shaped the spatial-temporal spread of COVID-19. A considerable research effort has been devoted to uncovering the dynamic interactions within this framework across different regions and pandemic stages using diverse geospatial big data (Wellenius et al., 2021). This study focused on published, peer-reviewed journal articles retrieved from the Web of Science search engine1. Web of Science is a subscription-based citation indexing service originally developed by the Institute for Scientific Information (ISI) and currently maintained by Clarivate Analytics. The platform is commonly used to access a vast collection of scientific literature and research papers with features including citation searches and analysis, author and institutional profiling, and research trend identification.

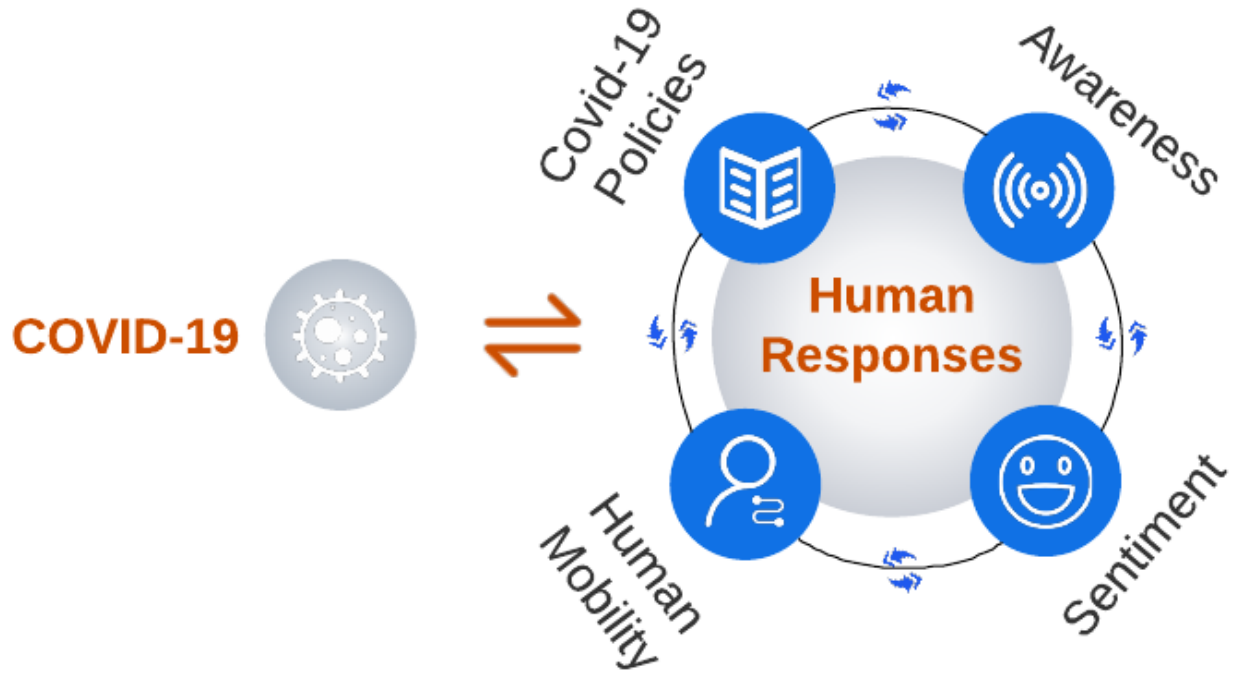

Figure 1. Conceptual framework of the study

The literature retrieval and analysis contain four steps. First, we used a specific procedure to retrieve the literature most relevant to human-COVID-19 interactions systematically. To manage the potentially large number of articles that a topic search could produce, we opted for a keyword-based search restricted to article titles. This allowed us to focus on the most pertinent literature on the subject. We used three categories of keywords, namely COVID-19, human responses, and relationship, as shown in Table 1. To be included in the search, the article's title had to contain at least one keyword from each category. The search was limited to articles published between 2020 and January 9th, 2023, and only publications in English with the document type as "article" were considered. To avoid retrieving irrelevant articles, we excluded

1. https://www.webofscience.com/wos/woscc/basic-search

several research areas, such as Chemistry, Microbiology, and Meteorology Atmospheric Sciences. This resulted in a total of 890 articles (Figure 2).

Secondly, we removed duplicate or irrelevant articles from the initial collection by manually reviewing the title and abstract of each article. A total of 572 articles were excluded in this step. Next, we conducted a narrative synthesis by manually reviewing the full text of each article to select studies involving geospatial data. This process resulted in 67 most relevant articles for the review.

Table 1. Keywords used in the initial collection of literature on human-pandemic interactions

| **Category** | **Keywords** |
|---|---|
| COVID-19 | COVID-19, pandemic, epidemic, coronavirus, SARS-CoV-2 |
| Human responses | Nonpharmaceutical Interventions, NPI, policy, policies, mobility, awareness, search engine, Google search, Baidu search, emotion, sentiment, attitude |
| Relationship | relationship, correlation, association, effect, impact, predict |

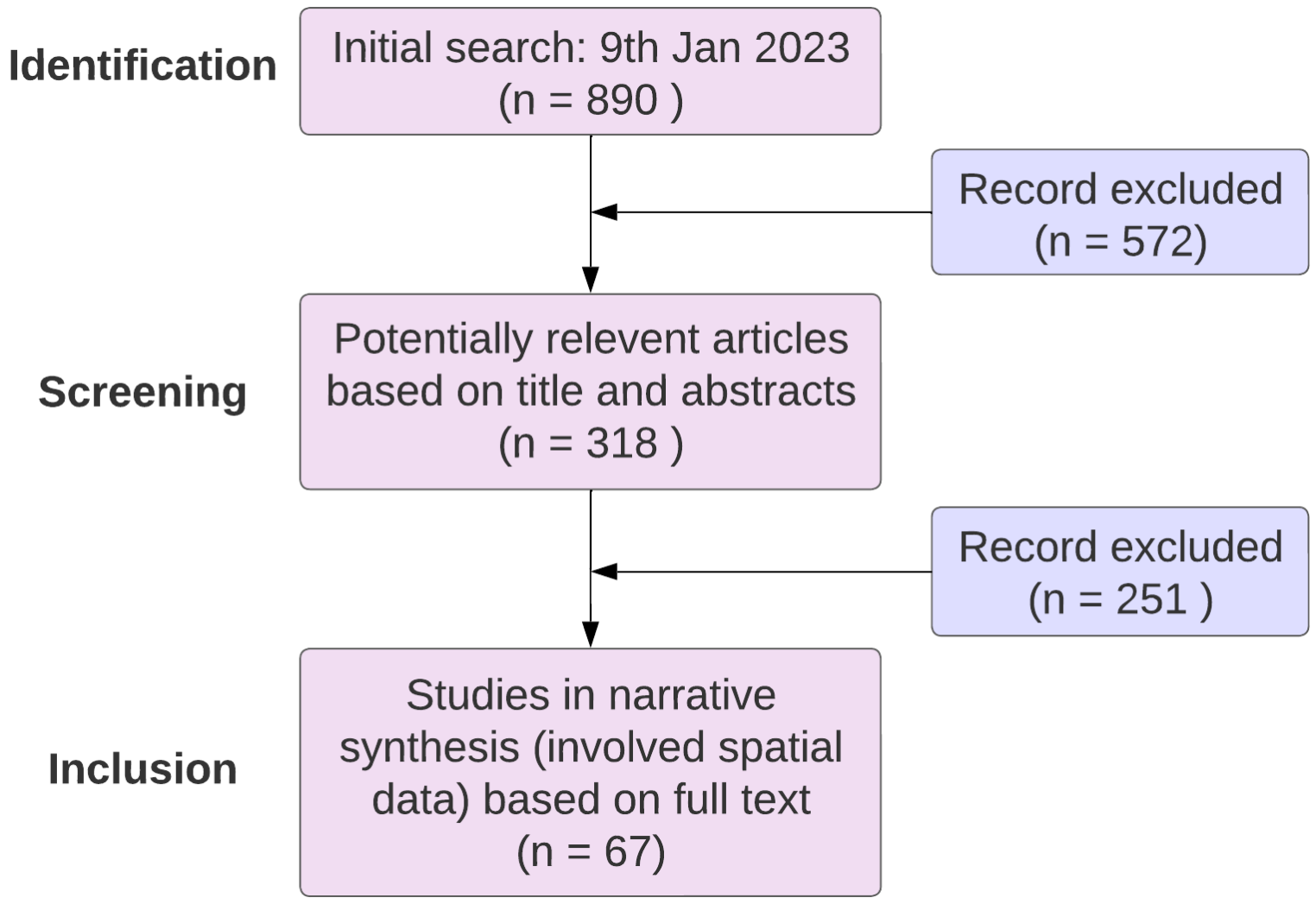

Figure 2. Inclusion criteria

Third, we created a review table to extract and record key information from each article to facilitate systematic content analysis and future ontological framework development. Each article was assigned a unique ID number. The review table consisted of six categories of information: (a) publication details, (b) research context, (c) research area and time, (d) data, (e) methodological framework, and (f) results and conclusions. Table 2 provides a detailed list of items under each category.

Table 2. Review Table

| **Category** | **Information Item** |
|---|---|
| Publication information | ID<br>Authors<br>Publication year<br>Paper title<br>First Author's affiliation<br>First Author's country<br>Journal name<br>Number of citations as of Jan 2023 (Web of Science) |
| Research context | Abstract<br>Keywords<br>Research objectives |
| Research area and time | Study area<br>Country<br>Geographic scale<br>Time period<br>Temporal scale |
| Data | Data name<br>Data source |
| Methodological framework | Human response variables and indicators<br>Pandemic health impact indicators<br>Model name<br>Model Source (<br>▪ Existing model,<br>▪ Improved models,<br>▪ New model) |
| Results and conclusion | Types of relationships (<br>▪ Spatially varying or unified,<br>▪ Dynamic or statistic,<br>▪ Causal or non-causal,<br>▪ Compounding or single chain,<br>▪ Bidirectional or one-directional)<br>Scope of relationships (<br>▪ The relationships among human responses,<br>▪ The effects of COVID-19 on human,<br>▪ The impacts of Human responses on COVID-19,<br>▪ The bidirectional relationships between COVID- |
|  | 19 and human responses) |
|  | Relationships |
|  | Other findings |
|  | Conclusion |

Utilizing data extracted from the review tables of these 67 articles, we conducted five analyses: (1) a statistical summary concerning publication information; (2) a summary of indicators pertaining to human responses and COVID-19; (3) delineating the scopes of relationship analysis; (4) enumerating the models employed by these articles to discern relationships among human-COVID-19 dynamics; and (5) encapsulating the findings regarding relationships over space and time from various perspectives.

## 3. Results

### *3.1 Overall Summary*

We gathered and organized publication data to obtain insights into three key aspects: the publication year, the country of origin of the first author, and the journals where the research was published (Figure 3). All articles analyzed in our study were published between 2020 and 2022, with 9, 31, and 27 papers published in 2020, 2021, and 2022, respectively. Notably, 26.87% of the first authors were based in the United States, while 16.42% were affiliated with institutions in China. Among the journals in which research within the field of human-COVID-19 dynamics were published, the *International Journal of Environmental Research and Public Health, PLOS ONE*, *Scientific Reports*, and *BMC Public Health* emerged as the top four publishers, accounting for approximately 8.96%, 7.46%, 5.97%, and 5.97% of the 67 papers.

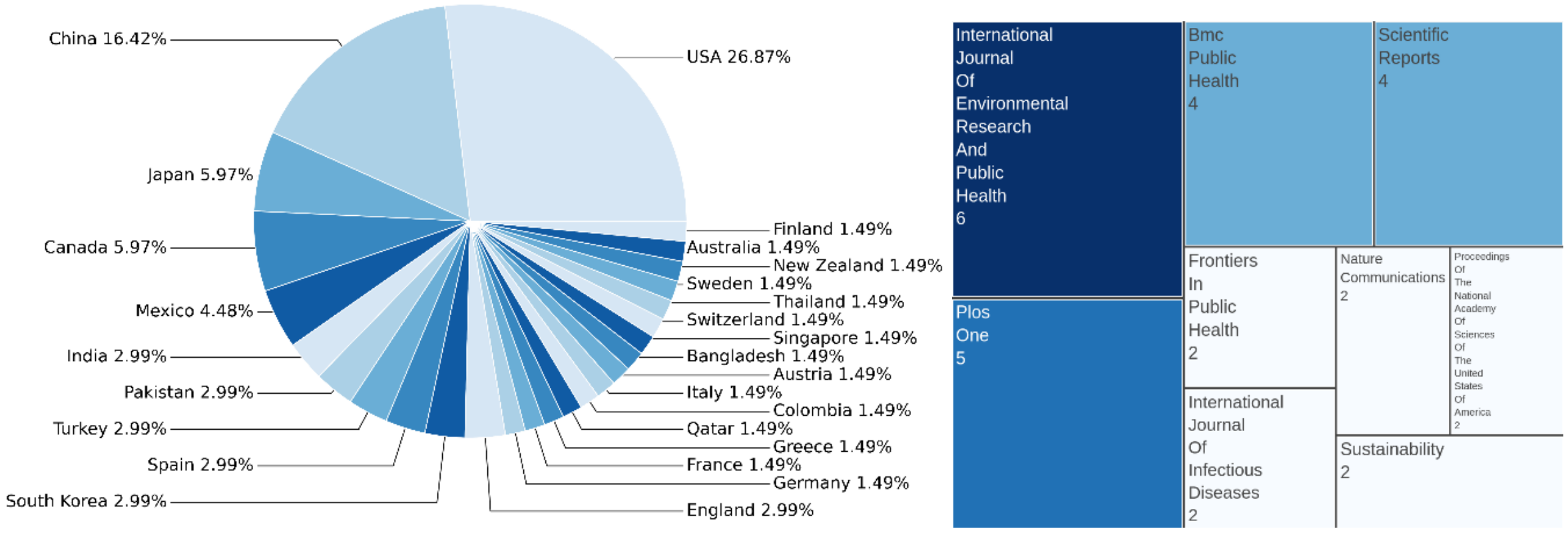

Figure 3. An overview of the countries of first authors' affiliations and the primary journal distribution.

Figure 4 presents an overview of the study area selection at the national level and the temporal scope of the study. In human-COVID-19 dynamics investigations, 70 countries were chosen as the focal points of analysis. United States emerged as the most frequently studied region, featuring in 26 (38.8%) papers. Italy, France, and China were selected as the study areas in 12 (17.9%), 11 (16.4%), and 11 (16.4%) papers, respectively. Conversely, many countries in Africa, Eastern Europe, Western Asia, and Central Asia were conspicuously absent as study areas within this research domain, possibility due to data unavailability. Given the spatial variability of human-COVID-19 dynamics, future efforts could aim to explore human-pandemic dynamics in these under-resourced regions to inform local pandemic control strategies. Regarding the temporal dimension, the majority of studies (47/70.1%) adopted a duration of no more than six months (Figure 4b). Approximately 22.4% of the papers analyzed data over a time span from six months to one year, while 7.5% of the papers extended their investigation to one year or longer.

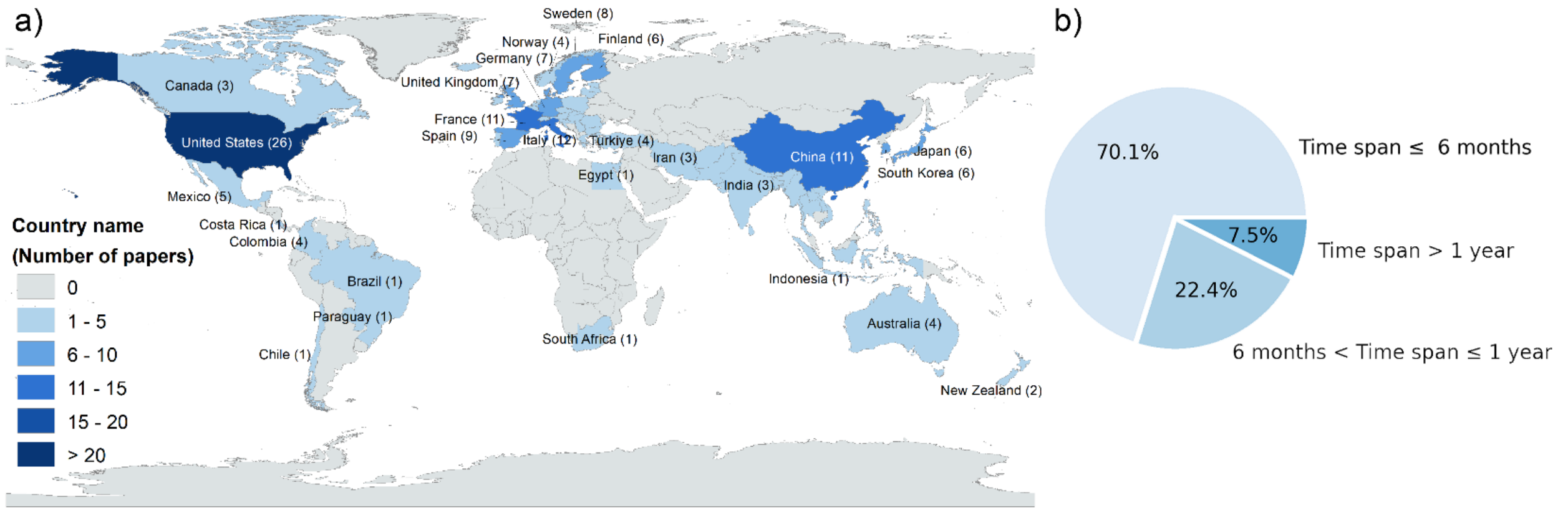

Figure 4. Summary of study area at country level and study time span.

Table 3 compiles the highest-cited articles within the reviewed collection, with citation counts retrieved from Web of Science as of March 2024. The range of citations spans from 1641 to 48. The leading article, published in Science, employed real-time travel history data from Wuhan, China, to elucidate the role of case importation in transmission across various Chinese cities. The second-ranked study, featured in Nature and being cited 695 times, empirically evaluates the impact of large-scale anti-contagion policies on the growth rate of infections.

Table 3. Ten most cited articles in the current collection.

| **Publication year** | **Paper title** | **Authors** | **Publication name** | **Number of citations as of March 2024** |
|---|---|---|---|---|
| 2020 | The effect of human mobility and control measures on the COVID-19 epidemic in China | Kraemer, Moritz U. G.; Yang, Chia-Hung; Gutierrez, Bernardo, et al. | Science | 1641 |
| 2020 | The effect of large-scale anti-contagion policies on the COVID-19 pandemic | Hsiang, Solomon; Allen, Daniel; Annan-Phan, Sebastien, et al. | Nature | 695 |
| 2020 | Association between mobility patterns and COVID-19 transmission in the USA: a mathematical modelling study | Badr, Hamada S.; Du, Hongru; Marshall, Maximilian, et al. | Lancet Infectious Diseases | 467 |
| 2020 | Effects of the COVID-19 Pandemic and Nationwide Lockdown on Trust, Attitudes Toward Government, and Well-Being | Sibley, Chris G.; Greaves, Lara M.; Satherley, Nicole, et al. | American Psychologist | 458 |
| 2020 | Association of the COVID-19 pandemic with Internet Search Volumes: A Google Trends$^{TM}$ Analysis | Effenberger, Maria; Kronbichler, Andreas; Shin, Jae Il, et al. | International Journal of Infectious Diseases | 143 |
| 2020 | Mobile device data reveal the dynamics in a positive relationship between human mobility and COVID-19 infections | Xiong, Chenfeng; Hu, Songhua; Yang, Mofeng, et al. | Proceedings of The National Academy of Sciences | 181 |
| 2020 | Analysis of mobility trends during the COVID-19 coronavirus pandemic: Exploring the impacts on global aviation and travel in selected cities | Abu-Rayash, Azzam; Dincer, Ibrahim | Energy Research & Social Science | 119 |
| 2021 | Associations of risk perception of COVID-19 with emotion and mental health during the pandemic | Han, Qing; Zheng, Bang; Agostini, Maximilian, et al. | Journal of Affective Disorders | 84 |
| 2021 | Impacts of social distancing policies on mobility and COVID-19 case growth in the US | Wellenius, Gregory A.; Vispute, Swapnil; Espinosa, Valeria, et al. | Nature Communications | 82 |
| 2021 | The Impact of Policy Measures on Human Mobility, COVID-19 Cases, and Mortality in the US: A Spatiotemporal Perspective | Li, Yun; Li, Moming; Rice, Megan; Zhang, Haoyuan, et al. | International Journal of Environmental Research And Public Health | 48 |

In Figure 5, the word cloud was derived from the abstracts of all 67 articles with text data preprocessing including stemming and lemmatization. The words of higher frequency are displayed in a larger font. As expected, the most frequently appearing words are related to disease, covid19, coronavirus, and outbreak. The middle-sized font words represent the second-highest frequency of use, and include terms such as behavioral, mitigate, prevent, intervention, and clinical. The smaller-sized font words include human, Wuhan, prevent, real-time, mobility, restriction, and population, offering additional insights into the thematic overview.

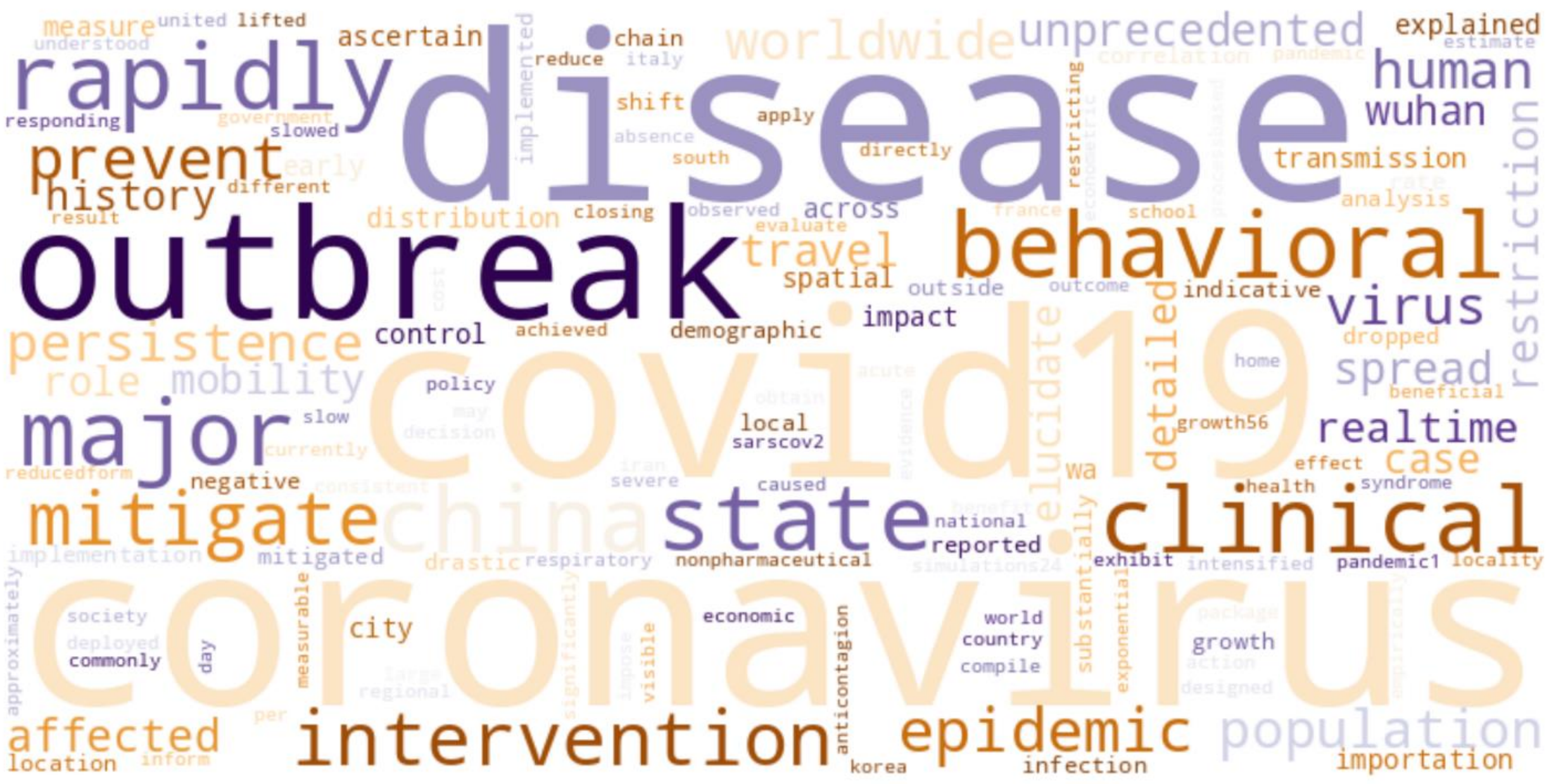

Figure 5. Word cloud derived from the abstracts of the reviewed 67 articles.

### *3.2 Indicators of human responses and COVID-19*

To quantitatively assess various aspects of human responses to COVID-19 for modeling human-COVID-19 dynamics, a range of indicators were employed, utilizing data of different spatial-temporal granularities. Figure 6 provides the distribution of articles that collected data across diverse geographical and temporal scales. Most articles (49.3%) gathered data at the national level, followed by 22.4% at the state or provincial level, 14.9% at the county level, and 9% at the city level. Regarding temporal scales, most articles (74.6%) collected daily data, while 14.9% of them focused on weekly data.

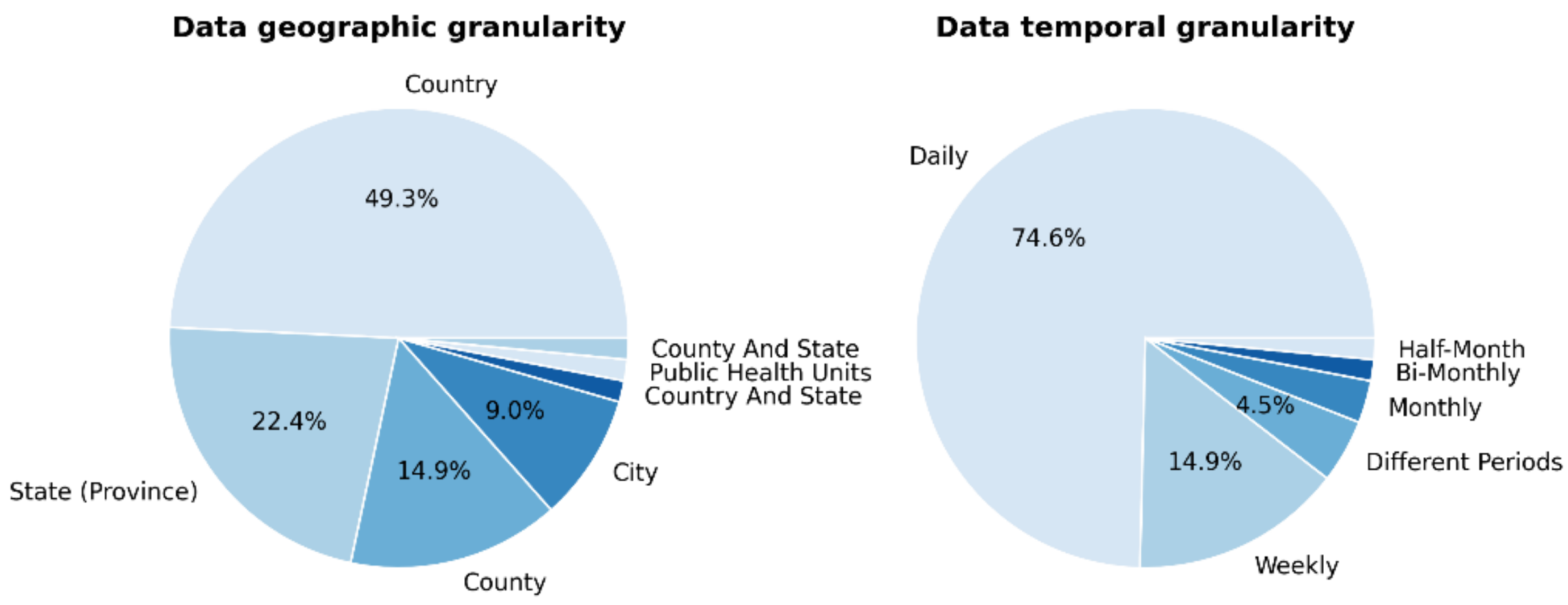

Figure 6. Spatial-temporal granularity of data.

A total of 52 papers considered the health impacts of COVID-19. These studies utilized various health impact indicators, including raw and population-normalized daily (cumulative) confirmed cases or deaths, the reproduction number (Rt), doubling time, infection growth rate, incidence rate ratios, and the logarithm of total COVID-19 cases. They were derived from diverse data sources: 10 papers (19.23%) relied on data from Johns Hopkins University's COVID-19 Dashboard, while 8 papers (15.38%) utilized data from official reports issued by provincial, municipal, or national health authorities. Additional data resources included the WHO Coronavirus Disease (COVID-19) Dashboard, Our World in Data, et al.

Table 4. Summary of COVID-19 health impacts data sources (proportion among total papers analysing COVID-19 health impacts).

| **Data Sources** | **Number of papers** | **References** |
|---|---|---|
| Johns Hopkins University's COVID-19 Dashboard | 10 (19.23%) | Badr et al., 2020; Buesa et al., 2021; Chang et al., 2021; Elitzur et al., 2021; He et al., 2021; Hsiang et al., 2020; Khan et al., 2021; Tu et al., 2021; Wellenius et al., 2021; Xiong et al., 2020 |
| Official reports from provincial, municipal, or national health governments | 8 (15.38%) | Dainton & Hay, 2021; W. Guo et al., 2022; Jung et al., 2021; Kraemer et al., 2020; Paternina-Caicedo et al., 2022; Poppe & Maskileyson, 2022; Vega-Villalobos et al., 2022; Wang et al., 2020 |
| WHO Coronavirus Disease COVID-19 Dashboard | 4 (7.69%) | P.-C. Chung & Chan, 2021; Cinarka et al., 2021; Tu et al., 2021; Wang et al., 2021 |
| Our World in Data | 4 (7.69%) | Rahman & Thill, 2022; Sözen et al., 2022; Tao et al., 2022; Widyasari et al., 2022 |
| The New York Times | 4 (7.69%) | Gottumukkala et al., 2021; Y. Guo et al., 2021; Kaufman et al., 2021; Zeng et al., 2021 |
| Control and Prevention (CDC) | 4 (7.69%) | Abbas et al., 2021; H. W. Chung et al., 2021; Effenberger et al., 2020; Jun et al., 2021 |
| Oxford COVID-19 Government Response Tracker (OxCGRT) | 3 (5.77%) | Bouzouina et al., 2022; C. W. S. Chen et al., 2022; de la Rosa et al., 2022 |
| Others | 15 (28.85%) | e.g., GitHub (Li et al., 2021), USAFacts (Yang et al., 2021), and China Data Lab (Y. Chen et al., 2021) |

Table 5 outlines the indicators and data sources related to COVID-19 policies, as referenced in a total of 38 papers. The Oxford COVID-19 Government Response Tracker (OxCGRT) is the primary data source for COVID-19 policies and has been utilized in 17 papers (44.74%). The stringency index, one of the OxCGRT indexes, offers a composite strictness measurement of COVID-19 policies based on nine response indicators, including five health system policy indicators (H1–H5, e.g., the COVID-19 testing regime and emergency healthcare investments) and four economic policy indicators (E1–E4, e.g., income support for citizens and foreign aid provision). The stringency index scales from 0 to 100 with 100 indicating the strictest measures. Other policy data sources encompass government websites, news websites, et al. These sources were employed to calculate composite policy indexes, policy implementation dates, and effective areas for quantitatively assessing COVID-19 policies.

Table 5. Summary of indicators and data sources of COVID-19 related policies (proportion among total papers analysing COVID-19 related policies).

| Indicators | Data Sources | Number of papers | References |
|---|---|---|---|
| • Stringency Index<br>• C (containment) indexes (C1-C8)<br>• H (health system policies) indexes (H1-H5)<br>• E (economic) indexes (E1-E4)<br>• Combination of C, H, E sindexes | Oxford COVID-19 Government Response Tracker (OxCGRT) | 17 (44.74%) | Bouzouina et al., 2022;<br>C. W. S. Chen et al., 2022;<br>H. W. Chung et al., 2021;<br>P.-C. Chung & Chan, 2021;<br>de la Rosa et al., 2022;<br>Gordon et al., 2021;<br>Kallidoni et al., 2022;<br>Khan et al., 2021;<br>Kumar et al., 2022;<br>Kuster & Overgaard, 2021;<br>Li et al., 2021;<br>Rahman & Thill, 2022;<br>Sözen et al., 2022;<br>Sukhwal & Kankanhalli, 2022;<br>Tao et al., 2022;<br>Wang et al., 2021;<br>Wu & Shimizu, 2022 |
| • Composite policy index<br>• Policy implementation date<br>• Policy implementation area<br>• Others | Government websites and various news websites | 10 (26.32%) | Chang et al., 2021;<br>Díaz-Castro et al., 2021;<br>Y. Guo et al., 2021;<br>Hsiang et al., 2020;<br>Jun et al., 2021;<br>Nguyen et al., 2021;<br>Poppe & Maskileyson, 2022;<br>Wellenius et al., 2021;<br>Yang et al., 2021;<br>Zhang et al., 2021 |

| | | |
|---|---|---|
| Others | 11 (28.95%) | e.g., Boston University School of Public Health COVID-19 US State Policy Database (Kaufman et al., 2021), the Global Database of Events, Language, and Tone (GDELT) (Gong et al., 2022), and the Institute of Health Metrics and Evaluation (IHME) (Elitzur et al., 2021) |

A summary of human mobility indicators and data sources concluded from 44 papers is presented in Table 6. The majority of human mobility indices were derived from locational data collected by map applications such as Google Maps, Apple Maps, and Baidu Maps (a map service widely used in China). The Google Mobility Indexes from Google's COVID-19 Community Mobility Reports are the most frequently employed, featuring in 21 papers (47.73%). The Google Mobility Indexes comprise users' direction requests volume change compared to the baseline across six location types, i.e., grocery and pharmacy, transit stations, retail and recreation, residential, parks, and workplaces. The Apple Mobility Report offers human mobility indices categorized by different travel modes, including transit, walking, and driving. These Apple mobility indices were used in 3 papers (6.82%). Migration Scale Index and Inter-city and Intra-city Mobility Indices from Baidu were applied in 2 papers (4.55%). Twitter, a popular social media platform, provides location tagging for users' tweets, enabling researchers to track the movement of users and analyze their geographical mobility. In this collection of papers, two indices, Cross-border Mobility and Daily Weighted Mobility Inflow received by county, were constructed using Twitter data. Additional datasets assessing human mobility encompass Teralytics (mobile phone tracking data in Switzerland), mobile terminal network operational data (Wu & Shimizu, 2022), AccuTracking software on cell phones (Pfeiffer et al., 2022), and taxi-trip datasets of Chicago (Mukherjee & Jain, 2022).

Table 6. Summary of indicators and data sources of human mobility (proportion among total papers analysing human mobility).

| **Indicators** | **Data Sources** | **Number of papers** | **References** |
|---|---|---|---|
| • Six Google mobility indexes (grocery and pharmacy, transit stations, retail and recreation, residential, parks, and workplaces)<br>• Average of six Google mobility indexes or five Google mobility indexes (exclude the residential)<br>• Partial Google mobility indexes by dimensionality reduction method, e.g., PCA | Google's COVID-19 Community Mobility Reports | 21 (47.73%) | Abulibdeh & Mansour, 2022;<br>Bouzouina et al., 2022;<br>Bryant & Elofsson, 2020;<br>Dainton & Hay, 2021;<br>Devaraj & Patel, 2021;<br>Díaz-Castro et al., 2021;<br>Gong et al., 2022;<br>Y. Guo et al., 2021;<br>He et al., 2021;<br>Jewell et al., 2021;<br>Jung et al., 2021;<br>Kumar et al., 2022;<br>Li et al., 2021;<br>Méndez-Lizárraga et al., 2022;<br>Paternina-Caicedo et al., 2022;<br>Rahman & Thill, 2022;<br>Sözen et al., 2022;<br>Wang et al., 2021;<br>Wellenius et al., 2021;<br>Widyasari et al., 2022;<br>Zhang et al., 2021 |
| • Three mobility indexes (transit, walking, driving)<br>• Average of three Apple mobility indexes | Apple Mobility Report | 3 (6.82%) | P.-C. Chung & Chan, 2021;<br>Kallidoni et al., 2022;<br>Wang et al., 2020 |
| • Migration scale index<br>• Inter-city and intra-city mobility indices | Baidu Inc. | 2 (4.55%) | Y. Chen et al., 2021;<br>Kraemer et al., 2020 |
| Three mobility indexes (road, train, and plane) | Teralytics (Mobile phone tracking data in Switzerland) | 2 (4.55%) | Badr et al., 2020;<br>Vinceti et al., 2022 |
| • Cross-border mobility<br>• Daily weighted mobility inflow received by county | Twitter API | 2 (4.55%) | Wallin Aagesen et al., 2022;<br>Zeng et al., 2021 |
| • Community Activity Score (CAS)<br>• Social Distance Index (SDI)<br>• Citymapper Mobility Index<br>• Time out of home<br>• Median daily activity living space<br>• Travel distance<br>• Air travel passenger count<br>• Traffic volume<br>• Others | Others | 14 (31.82%) | e.g., Mobile terminal network operational data (Wu & Shimizu, 2022), AccuTracking software on cell phones (Pfeiffer et al., 2022), and taxi-trip datasets of Chicago (Mukherjee & Jain, 2022) |

Public perception of COVID-19 encompasses both public awareness and sentiment, traditionally assessed through survey data. However, with the rise of social media platforms, residents have begun sharing their observations and thoughts online, making it possible to gauge public perception from a digital perspective. During the pandemic, due to policies like social distancing and lockdowns, many people turned to search engines to seek information about the virus, shared pandemic-related experiences and information on social media and expressed their emotional changes during this period. Researchers can assess public awareness and sentiment towards COVID-19 not only through surveys but also by examining digital sources such as search trends related to the pandemic or data from social media (e.g., Twitter and Facebook). For public awareness, 6 out of 10 studies (60%) utilized relative Google search volume as an index to estimate public awareness. Two studies (20%) derived awareness indexes from surveys, namely a Likert scale-based Public Awareness Index and Time Spent in Reading or Searching for Pandemic-Related Information. A total of 8 studies considered public sentiment towards COVID-19, and surveys served as the primary data source in 4 (50.00%) of them. These surveys employed psychological distress assessments and gauged various emotions using Likert scales. Natural language processing techniques enabled the capture of sentiment from text content on social media platforms. Sentiment indexes and percentages of positive, neutral, and negative sentiment were common types of indexes derived from social media data. Google search trends for emotion-related keywords, the Anxiety, Depression, Hopelessness, and Helplessness Index, and total sentiment indexes from the Global Database of Events, Language, and Tone (GDELT) were also utilized as alternative methods for evaluating public sentiment.

Table 7. Summary of indicators and data sources of public perceptions toward COVID-19 (proportion among total papers analysing public awareness/sentiment).

| | Indicators | Data Sources | Number of papers | References |
|---|---|---|---|---|
| COVID-19 Awareness | Relative Search Volume | Google/Baidu search trends for COVID-19 related keywords | 6 (60.00%) | Abbas et al., 2021; Cinarka et al., 2021; Effenberger et al., 2020; Jun et al., 2021; Kumar et al., 2022; Tu et al., 2021 |
| | • Public awareness index with a Likert scale<br>• Time in reading or searching for pandemic-related information | Survey | 2 (20%) | Han et al., 2021; Peng et al., 2022 |
| | Public awareness based on the numbers of past epidemics and human incidence or COVID-19 trends | Others | 2 (20%) | e.g., Emergency Events Database (Buesa et al., 2021) and Risk perception of COVID-19 linearly associated with the daily confirmed cases (Jung et al., 2021). |
| COVID-19 Sentiment | • Psychological distress with a Likert scale<br>• Different types of emotions with a Likert scale | Survey | 4 (50.00%) | Devaraj & Patel, 2021; Han et al., 2021; Peng et al., 2022; Sibley et al., 2020 |
| | • Sentiment index (-1~1) | Twitter or Facebook | 2 (25.00%) | Razzaq et al., 2022; Sukhwal & Kankanhalli, 2022 |
| | • Percentages of positive, neutral, and negative sentiment | | | |
| | Relative Search Volume | Google search trends for emotion related keywords | 1 (12.50%) | de la Rosa et al., 2022 |
| | Anxiety, depression, hopelessness, and helplessness index, and total sentiment index | Global Database of Events, Language and Tone (GDELT) | 1 (12.50%) | Gong et al., 2022 |

### *3.3 Scopes of relationships analysis*

The influence of COVID-19 on human society is multidimensional, encompassing government responses, public perceptions, and shifts in public behaviors, et al. These human responses interact with each other, forming an intricate spatial-temporal network of dynamic changes. This network, in turn, exerts reciprocal influence on the progression of the pandemic. Consequently, the Human-COVID-19 system represents an exceedingly complex, interlinked, dynamic, and regionally heterogeneous system. Typically, scientific articles concentrate on specific elements within the system, with a focus on exploring particular relationships that address scientific inquiries related to Human-COVID-19 dynamics. This section summarizes the current progress in researching these relationships within the Human-COVID-19 dynamics system, highlighting existing research gaps that may serve as future directions for researchers.

In Figure 7, we present a breakdown of research pertaining to the relationships between COVID-19 health impacts and human responses when considered as a holistic entity, as well as the internal dynamics of human responses. The total number of papers listed in Figure 7 is more than 67 because several studies investigated two or more relationships. Researchers have shown the greatest interest in understanding the influence of human responses on pandemic development (34 articles, 51% among all 67 papers). Exploring the internal relationships within human responses ranked second, accounting for 17 articles (25%). There are 9 articles (13%) investigating associations between human responses and COVID-19 health impacts without specifying the causality. Seven articles (10%) focus on how COVID-19 health impacts affect human responses. Only three (4%) articles delve into the bidirectional interactions between human responses and the evolution of COVID-19.

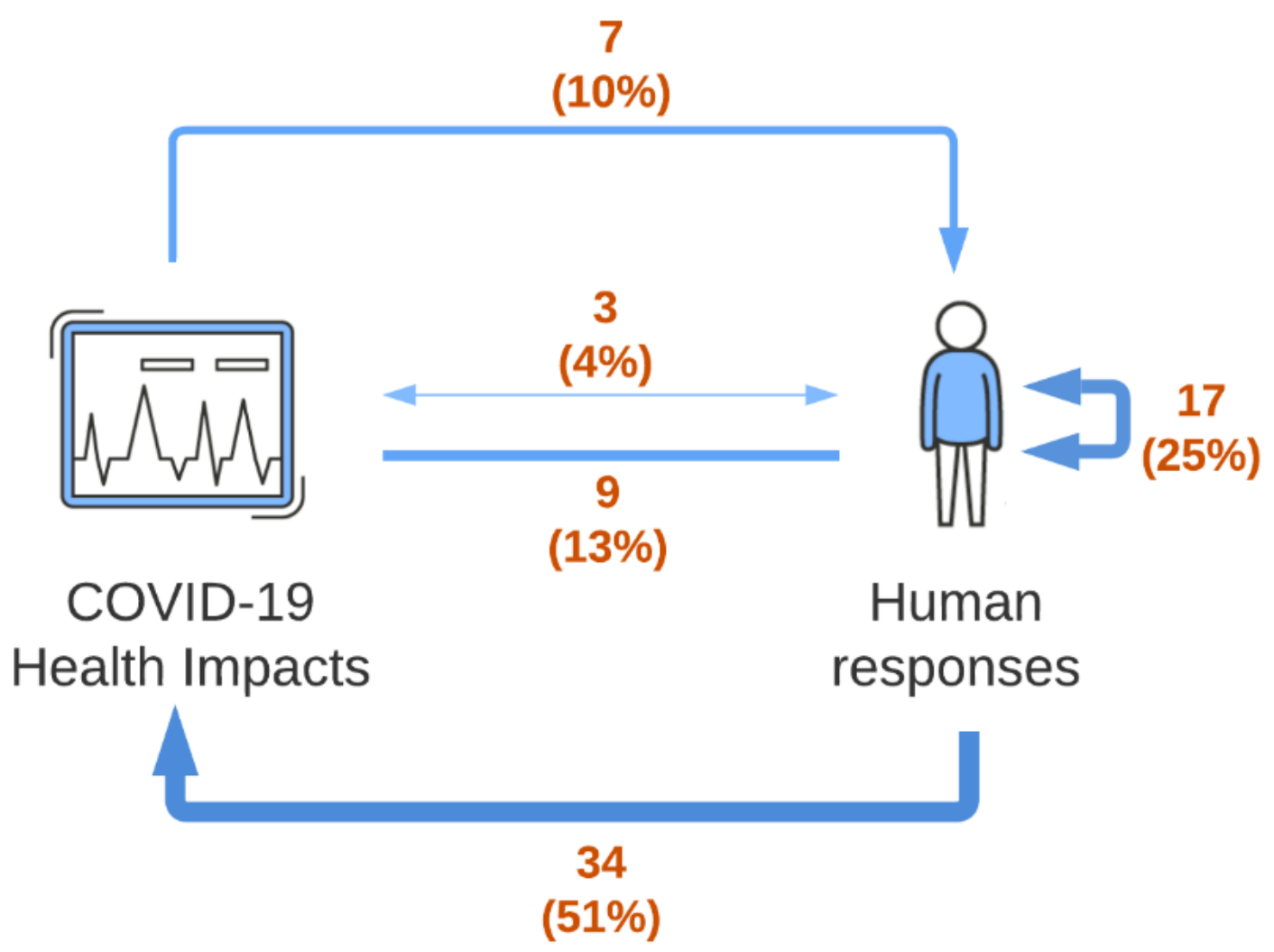

Figure 7. The amount and proportions of reviewed papers by relationship scopes.

Figure 8 provides a detailed breakdown of research discussing the relationships among variables within Human-COVID-19 dynamic systems. Studies correlate two variables are summarized on the left, and articles exploring relationships among more than two variables are outlined on the right. Among two-variable relationship investigations, research on the impact of COVID-19 policies on health impacts and research on the influence of human mobility on COVID-19 spread were the most prevalent, each comprising 12 articles. This suggests that policies and human mobility are the two factors that researchers are most concerned about in terms of their impact on the progression of the pandemic. The next popular research inquiry was the influence of policies on human mobility (8 articles). Additionally, 6 articles explored the direct effects of COVID-19 on human mobility, and 5 studies examined associations between COVID-19 and human mobility without specifying directions. Notably, there was a paucity of research on the roles of public sentiment and public awareness within the Human-COVID-19 system. Regarding studies involving multiple variables, 5 articles revealed the compounding influence of policies and human mobility on COVID-19 health impacts, while 4 articles delved into how COVID-19 policies affected human mobility, subsequently impacting COVID-19 spread.

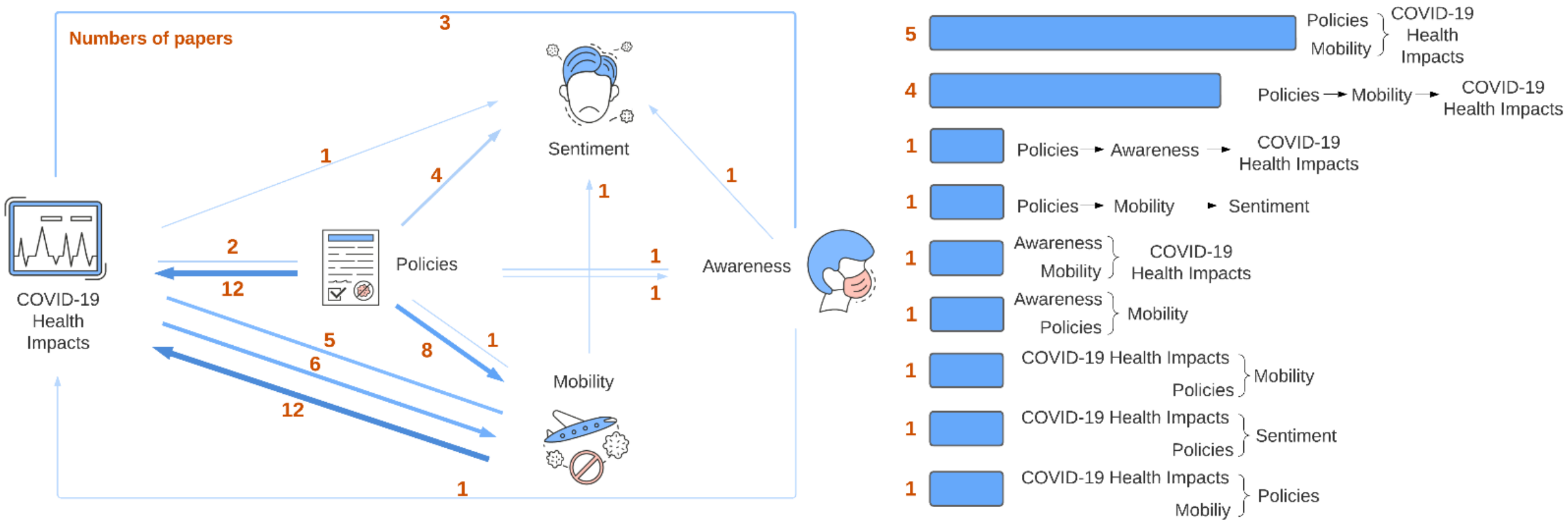

Figure 8. The amount and scope of papers analyzing two-variable relationships (left) and multi-variable relationships (right).

### *3.4 Relationship detection model*

A diverse array of models has been employed to discern relationships within the Human-COVID-19 dynamic systems. These models can be classified into four categories: causal (15 papers, 22.39%), correlation (13 papers, 19.40%), machine learning (3 papers, 4.48%), and regression (45 papers, 67.16%). Table 8 shows specific model names and the paper amount in the first three categories, while Table 9 furnishes in-depth information regarding regression models.

Table 8. An overview of models for relationship detection (proportion among all 67 papers).

| Model Type | Model name | Number of papers | References |
|---|---|---|---|
| Models for evaluating causal effects or intervention impacts | (1) Difference in Difference | 8 (11.94%) | Abu-Rayash & Dincer, 2020; Dainton & Hay, 2021; W. Guo et al., 2022; Nakamoto et al., 2022; Nguyen et al., 2021; Pfeiffer et al., 2022; Razzaq et al., 2022; Wallin Aagesen et al., 2022 |
| | (2) Interrupted time series | 3 (4.48%) | Chang et al., 2021; Kaufman et al., 2021; Poppe & Maskileyson, 2022 |
| | (3) Regression discontinuity approach | 2 (2.99%) | Sukhwal & Kankanhalli, 2022; Wellenius et al., 2021 |
| | (4) Multivariate analysis of variance (MANOVA) | 1 (1.49%) | Sibley et al., 2020 |
| | (5) Segmented regressions | 1 (1.49%) | Dainton & Hay, 2021 |
| Correlation-based | (1) Pearson correlation | 4 (5.97%) | Abulibdeh & Mansour, 2022; Gottumukkala et al., 2021; Sözen et al., 2022; Wu & Shimizu, 2022 |
| | (2) Time-lag correlation | 2 (2.99%) | Effenberger et al., 2020; Tu et al., 2021 |
| | (3) Cross Correlation | 1 (1.49%) | Vega-Villalobos et al., 2022 |
| | (4) Dynamic conditional correlation (DCC) | 1 (1.49%) | Cinarka et al., 2021 |
| | (5) Dynamic correlation | 1 (1.49%) | Abbas et al., 2021 |
| | (6) Functional canonical correlation analysis (FCCA) | 1 (1.49%) | Abbas et al., 2021 |
| | (7) Kendall’s coefficient of rank correlation | 1 (1.49%) | C. W. S. Chen et al., 2022 |
| | (8) Spearman’s rank correlation | 1 (1.49%) | C. W. S. Chen et al., 2022 |
| | (9) Sliding windows correlation models | 1 (1.49%) | Cinarka et al., 2021 |
| Machine learning | (1) XGBoost | 1 (1.49%) | Yang et al., 2021 |
| | (2) Random Forest | 1 (1.49%) | Tao et al., 2022 |
| | (3) Multilayer perceptron neural network algorithm | 1 (1.49%) | Sözen et al., 2022 |
| Other regression models | Details in Table 9 and 10 | 45 (67.16%) | Details in Table 9 and 10 |

Causal models operate under the assumption that, in the absence of intervention, the subject would follow parallel trends over time. Causal inference is drawn from observed diverging trends following the implementation of the intervention. As indicated in Table 8, the most frequently employed causal model was the *Difference in Difference* method, which was used in 8 papers (11.94%). Additionally, 3 papers (4.48%) utilized the *Interrupted Time Series model*, and 2 papers (2.99%) applied the *Regression Discontinuity approach*. *Multivariate Analysis of Variance (MANOVA)* and *Segmented Regressions* were each utilized once in the reviewed literature. While these models can provide valuable insights, they may not always yield truly causal estimates if the underlying assumptions are not met or if uncontrolled confounding factors exist. In correlation models, the Pearson correlation was applied in 4 papers (5.97%). Other models in this category included *Time-Lag Correlation*, *Cross Correlation*, et al.  Machine learning models, including complex architectures like deep neural networks, can capture intricate non-linear relationships within Human-COVID-19 dynamics. Among collected studies, 3 adopted different machine learning algorithms, namely, *XGBoost*, *Random Forest*, and *Multilayer Perceptron Neural Network*.

Table 9. Categories of regression models for relationship detection

| **Model Type** | | **Basic** | **Panel data regression model** | | **Time series model** | **Spatial model** | **Spatial-temporal model** |
|---|---|---|---|---|---|---|---|
| | | | **Type 1** | **Type 2** | | | |
| Input data | Time-varying ( √ ) or Static (×) | × | √ | √ | √ | × | √ |
| | Spatially varying ( √ ) or Unified (×) | × | √ | √ | × | √ | √ |
| Spatial dependence | Considered ( √ ) or not considered (×) | × | × | × | × | √ | √ |
| Output relationships | Time-varying ( √ ) or Static (×) | × | × | √ | √ | × | √ |
| | Spatially varying ( √ ) or Unified (×) | × | × | × | × | √ | √ |

Table 10. An overview of regression models in different categories (proportion among all 67 papers)

| Model Type | Model name | Number of papers | References |
| --- | --- | --- | --- |
| Basic | Simple linear regression | 8 (11.94%) | de la Rosa et al., 2022; Devaraj & Patel, 2021; Elitzur et al., 2021; Gordon et al., 2021; Jun et al., 2021; Kraemer et al., 2020; Lison et al., 2022; Tokey, 2021 |
| Panel data regression model_Type1 | (1) Generalized linear regression model | 2 (2.99%) | Badr et al., 2020; Wang et al., 2020 |
| | (2) Simultaneous equations model | 1 (1.49%) | Xiong et al., 2020 |
| | (3) Structural equation model | 1 (1.49%) | Rahman & Thill, 2022 |
| | (4) Bayesian multilevel generalized structural equation model | 1 (1.49%) | Zhang et al., 2021 |
| | (5) Partial least square Structural equation model | 1 (1.49%) | Widyasari et al., 2022 |
| | (6) Poisson regression model | 1 (1.49%) | H. W. Chung et al., 2021 |
| | (7) Log-linear regression model | 1 (1.49%) | Jung et al., 2021 |
| | (8) Reduced-form analysis | 1 (1.49%) | Hsiang et al., 2020 |
| Panel data regression model_Type2 | (1) Linear mixed-effects models (Multilevel linear regressions) | 4 (5.97%) | P.-C. Chung & Chan, 2021; Han et al., 2021; Jewell et al., 2021; Kuster & Overgaard, 2021 |
| | (2) Pooled Mean Group–Autoregressive Distributed Lag (PMG-ARDL) model | 2 (2.99%) | Bouzouina et al., 2022; Khan et al., 2021 |
| | (3) Fixed effects model | 2 (2.99%) | Gong et al., 2022; Li et al., 2021 |
| | (4) Panel Regression model | 2 (2.99%) | Díaz-Castro et al., 2021; Kumar et al., 2022 |
| | (5) Panel Vector Autoregression model | 1 (1.49%) | Wang et al., 2021 |
| | (6) Mixed-effects Poisson model | 1 (1.49%) | Méndez-Lizárraga et al., 2022 |
| | (7) Hierarchical regression analyses | 1 (1.49%) | Peng et al., 2022 |
| | (8) Negative Binomial regression model | 1 (1.49%) | Y. Guo et al., 2021 |
| Time series model | (1) Autoregressive integrated moving average (ARIMA) model | 3 (4.48%) | de la Rosa et al., 2022; Jun et al., 2021; |
| | | | Paternina-Caicedo et al., 2022 |
| | (2) Autoregressive model | 1 (1.49%) | Yang et al., 2021 |
| | (3) Poisson count time series | 1 (1.49%) | Zeng et al., 2021 |
| | (4) Markov-Chain Monte-Carlo (MCMC) model | 1 (1.49%) | Bryant & Elofsson, 2020 |
| | (5) Seasonal Autoregressive Integrated Moving Average with exogenous regressors (SARIMAX) time-series model | 1 (1.49%) | Kallidoni et al., 2022 |
| | (6) Seasonal Autoregressive Integrated Moving Average (SARIMA) intervention analysis model | 1 (1.49%) | Meng et al., 2021 |
| | (7) Newey-West linear regression model | 1 (1.49%) | Vinceti et al., 2022 |
| Spatial model | (1) Geographically Weighted Regression (GWR) | 1 (1.49%) | Tokey, 2021 |
| | (2) Spatial regressions model | 1 (1.49%) | Buesa et al., 2021 |
| | (3) Spatial error model | 1 (1.49%) | Tokey, 2021 |
| Spatial- temporal model | (1) Bayesian spatiotemporal generalized additive mixed model (GAMM) model | 1 (1.49%) | He et al., 2021 |
| | (2) Geographically and Temporally Weighted Regression (GTWR) | 1 (1.49%) | Y. Chen et al., 2021 |

Regression models play a predominant role in this research direction and were used in 45 (67.16%) papers. To offer a more comprehensive overview of regression models, we categorized them based on three key criteria: the inclusion of spatial-temporal information in the input data, consideration of spatial dependence within the model, and the potential for output relationships to vary over space and time (Table 9). As indicated in Table 10, the most foundational model is the simple linear regression, which was employed in 8 papers. Simple linear regression uses input data devoid of spatial-temporal information, does not incorporate spatial dependence within the model, and yields consistent results across time and space. The second category encompasses panel data regression models, which account for geographic and temporal variations in input data but do not explicitly incorporate spatial dependence into model structures. Consequently, output relationships remain constant regardless of geographical location. Within this category, we further subdivided these models based on whether output relationships change over time. The first subcategory includes models where output relationships remain static over time, such as the *Generalized Linear Regression Model, Simultaneous Equations Model*, et al. The second subcategory includes models where output relationships evolve with time, including *Linear Mixed-Effects Models*, *PMG-ARDL Model*, et al.

The third category is time series models, where both input data and output relationships vary over time. However, these models do not consider geographic heterogeneity. Models in this category include the *ARIMA* Model, *Autoregressive Model*, et al. Among these models, *ARIMA* was the most frequently employed, applied in 3 papers, while the others were utilized only once. The fourth category is spatial models, where both input data and output results vary across space, with consideration of spatial dependence. However, these models do not incorporate temporal variations. This category includes *GWR*, *Spatial Regressions Model*, and *Spatial Error Model*.

The fifth category in regression models is spatial-temporal models, which account for spatial-temporal heterogeneity, spatial dependence, and estimate relationships over space and time. Two papers employed this category, with one utilizing the *Bayesian GAMM* and the other employing the *GTWR*.

### ***3.5 Spatial-temporal relationships***

#### *3.5.1 The effects of COVID-19 policies on pandemic health impacts*

Government responses to mitigate the COVID-19 spread have been at the forefront of pandemic management efforts. Numerous studies have assessed the effectiveness of these policies in curbing the pandemic's impact. Hsiang et al. (2020) evaluated COVID-19 policies' impacts in China, South Korea, Italy, Iran, France, and the United States, using a reduced-form econometric method. In the absence of policy interventions, early COVID-19 infections exhibited an exponential growth rate of approximately 38% per day. Policy measures significantly reduced transmission rates, preventing an estimated 61 million confirmed cases.

Despite an overall positive assessment of COVID-19 policies' efficacy, varying policy types exhibited diverse effectiveness in mitigating health impacts. The top three policies associated to second wave growth global scale were mandatory facial coverings in public, limitations on gatherings of ten people or fewer, and screening of foreign travelers on international flights (Tao et al., 2022). Chung et al. (2021) categorized countries based on the number of pandemic waves and found that contact tracing and containment policies were effective in containing the pandemic for countries with two waves, while closure, economic, and health policies were useful for countries with three waves. In Nordic countries, early-stage restrictions on international travel effectively reduced COVID-19 cases during the first half of 2020 (Gordon et al., 2021). Social distancing measures were linked to a 15.4% daily reduction in COVID-19 cases, preventing nearly 33 million cases nationwide within three weeks in the U.S. (Kaufman et al., 2021). In the United States, school closures significantly reduced the basic reproductive number at the county level during the first half of 2020 (Yang et al., 2021).

In addition to policy types, the timing of policy implementation emerged as a critical factor affecting COVID-19 spread. Elitzur et al. (2021) found that the timing of policy implementation crucially impacts COVID-19 health outcomes based on data from 89 nations and US states. Stay-at-home policies lost effectiveness when implemented in the later phase of an outbreak. Delaying policy implementation by one week could nearly triple the infected population.

The effects of COVID-19 policies on the pandemic control varied in the short and long terms. Based on the analysis of South Asian countries from January 2020 to May 2021, economic support, stringency, and health and containment indices (an expanded index builds on the Stringency Index) effectively reduced the pandemic's impact in the long term. While, in the short term, only the health and containment index effectively reduced the risk of resurgence (Khan et al., 2021).

### *3.5.2 The impacts of COVID-19 policies on human mobility*

Since the outbreak of COVID-19 pandemic, governments worldwide have enacted diverse policies aimed at curbing virus transmission by regulating human movement. These measures include stay-at-home orders, school and workplace closures, travel restrictions, event cancellations, and public transport suspensions. Understanding how these policies impact human mobility is vital for effective pandemic control. Research aims to explore variations in policy effectiveness based on type and timing, regional disparities in impact, and potential time lags in policy effectiveness. Wellenius et al. (2021) found that in the U.S., state-level emergency declarations reduced overall mobility by 9.9%, with additional reductions of 24.5% with social distancing policies (34.4%), and 29.0% with shelter-in-place mandates (38.9%).  Kallidoni et al. (2022) discovered that school closures had the most significant impact on reducing driving and walking across twenty-five European countries from February 2020 to February 2021, while stay-at-home orders had limited effects on mobility reduction. However, in the U.S., stay-at-home orders proved effective in mobility reduction from March to mid-July 2020 (Li et al., 2021). Workplace closures were also linked to notable decreases in overall mobility, while public information campaigns had minimal influence. In South Korea, mobility in retail and recreation, transit stations, and residential region correlated with the stringency of COVID-19 stay-at-home polices, while visitations to grocery and pharmacy, parks, and workplaces showed no significant relationship with policies from March 2020 to February 2021 (Sözen et al., 2022).

The impact of pandemic policies on human mobility varies across regions and pandemic phases. Nakamoto et al. (2022) studied the effects of two emergency declarations in Japan during different stages of the COVID-19 pandemic from February 1, 2020, to April 30, 2021. They found that while both declarations reduced human mobility, the impact of the second emergency declarations was slightly weaker. This suggests that the efficacy of emergency declarations in reducing human mobility to control pandemic spread diminished over time. Chang et al. (2021) examined spatial heterogeneity in the effects of COVID-19 policies on human mobility in the U.S. at the county level. They concluded that socioeconomically disadvantaged counties were less affected by stay-at-home orders compared to counties with more resources.

Quantifying the possible delayed effects of policy on reducing human mobility is crucial for determining the optimal timing of policy enactment. The implementation of pandemic policies involves several steps, including proclamation, information dissemination, compliance adaptation, and eventual observable changes in behavior. This process often entails a temporal delay in the impact of pandemic policies on human mobility. Sözen et al. (2022) found no delay in the impact of COVID-19 stay-at-home policies in Poland, Turkey, and South Korea from March 2020 to February 2021. Conversely, Wellenius et al. (2021) observed a 24.5% mobility reduction occurred one week after the implementation of social distancing policies in the U.S. between January and March 2020. This observation was made through regression discontinuity analysis utilizing data from Google COVID-19 Community Mobility Reports.

*3.5.3 The relationships between human mobility change and COVID-19 spread*

Notable human mobility decrease has been discovered at multiple geographical scales after the COVID-19 outbreak, partially due to the implementation of relevant policies. Amidst the COVID-19 pandemic, human mobility decreased compared to the pre-COVID-19 period in China (W. Guo et al., 2022), United States (Pfeiffer et al., 2022), and Nordic countries (Wallin Aagesen et al., 2022). Wallin Aagesen et al. (2022) further demonstrates that decreases in cross-country-border mobility ranged from −35% (Iceland) to −82% (Finland). At the city level, Mukherjee & Jain (2022) analyzed urban mobility data in Chicago from March 2020 to November 2020, and found a sharp decline in travel-related demand in regions with high economic activities, e.g., airports, downtown areas, and business zones.

These human mobility reductions have subsequently influenced the spread of COVID-19. The cumulative number of cases outside Wuhan, China, was positively correlated with population inflow from Wuhan (Y. Chen et al., 2021). Xiong et al. (2020) highlighted a positive relationship between mobility inflow and the number of infections, which indicated the efficiency of limiting inflow mobility to mitigate pandemic spread. This phenomenon became increasingly stronger in partially reopened regions in the U.S. from March 1 to June 9, 2020. In the state capitals of Colombia, case rates decreased as mobility in retail stores reduced (Paternina-Caicedo et al., 2022). Jewell et al. (2021) found that in the U.S., the positive relationships between human mobility and COVID-19 cases were significant in Spring 2020, decreased in summer and fall 2020, and then increased in late 2020 and early 2021.

Some studies indicate varying delays in the positive impacts of reduced human mobility on controlling COVID-19 spread. For instance, New York and Madrid experienced a 14-day and 18-day delay, respectively, in the decrease in deaths correlating with the reduction of public transport mobility from March to October 2020 (Vega-Villalobos et al., 2022). A delay of approximately 3 weeks was observed in increased infections following mobility increases at the county level in the U.S. from January to April 2020 (Badr et al., 2020). Longer delays of 5-7 weeks in case number growth were observed following mobility growth patterns in 20 U.S. states from July to September 2020 (Gottumukkala et al., 2021), as well as at the county level in the U.S. from February to July 2020 (He et al., 2021).

Conversely, reductions in human mobility have been observed to correlate with increased transmission of COVID-19 in some studies. In Mexico City, although a reduction in the total number of daily trips taken via public transport was observed, the daily COVID-19 deaths increased from March to October 2020 (Vega-Villalobos et al., 2022). In the U.S., infection rates were negatively correlated with average travel miles per person and out-of-county trips nationwide from March to August 2020 (Tokey, 2021). A few other studies have found insignificant relationships between some types of human mobility and COVID-19 spread. For instance, visitation to parks was not related to COVID-19 spread in the U.S. at the county level from February to July 2020 (He et al., 2021). The visitation to Grocery was also found not related to the number of infections in European countries from 12 March 2020 to 31 August 2021(Bouzouina et al., 2022).

### *3.5.4 The associations between COVID-19 awareness and pandemic*

Public awareness of disasters or crises plays a critical role in shaping human behaviors in responding to disasters or crises. Previous work has attempted to estimate public awareness of COVID-19 and decipher its relationship with the pandemic spread. Cinarka et al. (2021) utilized Google search trends of COVID-19 symptoms to gauge public awareness in Turkey, Italy, Spain, France, and the United Kingdom from January 1 to August 31, 2020. The study revealed that symptoms like fever, cough, and dyspnea correlated well with new cases during the first wave, but correlations began fluctuating after May 2020. Effenberger et al. (2020) examined the link between public awareness and COVID-19 cases in different countries using Google search trends for "Coronavirus" compared to reported cases by the European Center for Disease Control (ECDC). Their analysis showed a consistent positive correlation, with peak interest occurring 11.5 days before the peak in reported cases, observed across European countries and the US. In Brazil and Australia, peak correlations occurred 7 days prior, while in Egypt, there was no lag. Tu et al. (2021) examined search trends for common COVID-19 symptoms on Baidu, a widely used search engine in China, from January 11 to April 22, 2020. Spearman’s correlation analysis revealed strong positive correlations between daily confirmed cases and Baidu search trends for each symptom. The average delay of increased confirmed cases after the search peak was 19.8 days. Specifically, confirmed cases lagged behind the "cough" search trend by 4 days, “fatigue” by 2 days, “sputum production” by 3 days, “shortness of breath” by 1 day, and “fever” by 0 days. Jung et al. (2021) used public awareness, human mobility, and temperature to predict the effective reproduction number of COVID-19 in Japan during January 16, 2020 to February 15, 2021. Their analysis suggested that the model including public awareness performed better than models without public awareness, and public awareness was negatively associated with COVID-19 transmission.

### *3.5.5 Influential factors in shaping COVID-19 sentiment*

The implementation of COVID-19 policies significantly influences public sentiment towards the virus. During the nationwide lockdown in New Zealand, residents experienced higher rates of mental distress compared to the pre-lockdown phase, as indicated by the New Zealand Attitudes and Values Study (Sibley et al., 2020). In Singapore, public sentiment evolved over time, with an increase during the lockdown period, a further rise after partial lifting of restrictions, and a subsequent decrease following further easing of measures (Sukhwal & Kankanhalli, 2022). In Shenzhen, populations in lockdown areas exhibited more negative social emotions compared to other regions, as revealed by a survey examining social emotions conducted across four zones, i.e., lockdown, control, prevention, and safe zones, during a week-long lockdown from March 13 to March 20, 2020 (Peng et al., 2022).

The COVID-19 health impacts were also found to be a determinant of public sentiment. Through analyzing the "Quarantine Life" dataset, including thousands of tweets from India, Ireland, Midrand, the United States, and South Africa from January to September 2020, Razzaq et al. (2022) revealed that individuals experienced distress and fear during the COVID-19 pandemic. COVID-19 sentiment was closely linked to COVID-19 awareness. A significant link was observed between higher risk perception of COVID-19 and more negative public sentiment in the PsyCorona Survey which involved 54,845 participants across 112 countries (Razzaq et al., 2022). Human mobility serves as another determinant of public sentiment. Devaraj & Patel (2021) investigated how psychological distress changed in response to reduced mobility during the early stages of the 2020 COVID-19 pandemic in the United States. They analyzed data from 5,132 individuals who participated in the March and April 2020 waves of the Understanding America Study (UAS) and found that a one standard deviation decline in mobility was associated with a 3.02% higher psychological distress.

## 4. Discussion

This synthesis analysis presents a comprehensive overview of past efforts to investigate the relationships among human-pandemic dynamic systems using geospatial big data. Researchers have collectively devoted great efforts in discovering the mechanism within these systems, as evidenced by the diverse study objects, data sources, indicators of human responses and COVID-19 health impacts, scopes of relationships analysis, relationship detection models, and numerous insightful findings. At the same time, it points out existing challenges across different domains that necessitate further research attention to bridge knowledge gaps.

First, the multitude of human responses to COVID-19 are interconnected factors that give rise to numerous closed influencing loops. For example, prolonged adherence to human mobility restrictions may induce fatigue and reduce compliance with safety measures and potentially foster negative sentiments towards the pandemic. Diminished awareness and negative sentiments might prompt individuals to be less compliant with stay-at-home policies, thereby increasing human mobility. Despite these interconnected dynamics, there is a noticeable dearth of research attention dedicated to exploring these feedback loops among human responses. Moreover, the emergence of COVID-19 has instigated interactions among human reactions, which, in turn, impact the spread of the virus. Understanding these reciprocal relationships is crucial for comprehending the pandemic’s effects on humanity and grasping effective pandemic control. However, our findings reveal a scarcity, with only three articles (4%) delving into bidirectional interactions between human responses and the evolution of COVID-19. Future research should expand the scopes of relationship analysis, adopting a more comprehensive perspective and conducting analyses of the interconnected, closed-loop influencing, bidirectional relationships among various elements within human-pandemic dynamics, enabling more efficient and effective pandemic response efforts in the future.

Second, the inclusion of COVID-19 awareness and sentiment in human-pandemic dynamic analyses is limited, with only 10 articles (15%) and 8 articles (12%) incorporating these factors. Awareness and sentiment play a significant role in shaping individual behavior and decision-making regarding preventive measures and adherence to public health guidelines, thus directly influencing the spread of the pandemic. Moreover, comprehending how individuals perceive and respond to the pandemic on a psychological level is crucial for developing interventions that not only effectively and sustainably curb the spread of the virus but also embody humanistic care.

Third, despite 15 papers (22.39%) identifying causal relationships utilizing classical statistical models such as Difference in Difference and Interrupted Time Series, there exists an imperative for further exploration. Causal modeling entails the consideration and control of potential confounding factors, facilitating the inference and prediction of causal relationships between specific variables rather than mere correlations. Such models, compared to association models, are capable of unveiling deeper underlying mechanisms, evaluating the effects of specific interventions on COVID-19 spread, and more precisely forecasting pandemics. This ensures that we can derive more accurate and reliable conclusions, thereby enhancing our comprehension of the intricate causal dynamics inherent in human-pandemic interactions.

Fourth, only two papers (2.99%) employ spatial-temporal models, which emerge as the most appropriate for analyzing data imbued with inherent spatial-temporal information in human responses and COVID-19 health impacts. Incorporating spatial-temporal heterogeneity and spatial dependence into research methodologies enables a more sophisticated understanding of how human activities and COVID-19 spread vary and interact across regions and time periods. The ability to estimate spatial-temporally sensitive relationships enhances our predictive capabilities and responsiveness to COVID-19. Consequently, prioritizing the development and implementation of advanced spatial-temporal models is imperative.

Finally, our findings reveal that only 5 papers (7.46%) set the interest time window to one year or longer, despite the enduring nature of the pandemic. The choice of the research time frame significantly influences analytical outcomes, with longer temporal spans providing a more accurate portrayal of the temporal scale and yielding results closer to reality. Given that we have generated and collected sufficient data over the past four years since the beginning of the pandemic, extending the study period beyond two years is both feasible and necessary.

## 5. Conclusions

This synthesis study aimed to deepen the understanding of human-pandemic dynamics using geospatial big data by analyzing 67 selected journal articles from March 25th,2020 to January 9th, 2023. Our findings reveal that various forms of geospatial big data were utilized in studying human-COVID-19 interactions, including location-based social media data, website data, and location-based usage/log data. Among the selected literature, 52, 44, 38, 10, and 8 articles considered the COVID-19 health impacts, human mobility, COVID-19 policies, public awareness, and public sentiment, respectively.  Regression models were the most popular approach for detecting human-pandemic relationships (45 papers), but only two studies leveraged spatial-temporal models. Research on the impact of policies on COVID-19 health impacts and research on the influence of human mobility on COVID-19 spread were the most prevalent, each comprising 12 articles. Only 3 papers delved into the bidirectional interactions between human responses and the evolution of COVID-19. Our examination of human-pandemic dynamics highlighted five key aspects, including the effects of COVID-19 policies on health impacts, the impacts of policies on human mobility, relationships between changes in human mobility and COVID-19 spread, associations between COVID-19 awareness and the pandemic, and triggers of COVID-19 sentiment.

While our study focused on a limited timeframe, it has provided valuable insights into future directions for research on human-pandemic dynamics. The directions are as follows: extending the temporal scope of investigation beyond two years; broadening the scope of relationship analysis among more elements; delving into the examination of causal relationships; developing spatial-temporal modeling; and analyzing the interconnected, closed-loop, bidirectional relationships within human-pandemic dynamics.

## References

Abbas, M., Morland, T. B., Hall, E. S., & EL-Manzalawy, Y. (2021). Associations between Google Search Trends for Symptoms and COVID-19 Confirmed and Death Cases in the United States. *International Journal of Environmental Research and Public Health*, *18*(9), Article 9. https://doi.org/10.3390/ijerph18094560

Abulibdeh, A., & Mansour, S. (2022). Assessment of the Effects of Human Mobility Restrictions on COVID-19 Prevalence in the Global South. *The Professional Geographer*, *74*(1), 16–30. https://doi.org/10.1080/00330124.2021.1970592

Abu-Rayash, A., & Dincer, I. (2020). Analysis of mobility trends during the COVID-19 coronavirus pandemic: Exploring the impacts on global aviation and travel in selected cities. *Energy Research & Social Science*, *68*, 101693. https://doi.org/10.1016/j.erss.2020.101693

Badr, H. S., Du, H., Marshall, M., Dong, E., Squire, M. M., & Gardner, L. M. (2020). Association between mobility patterns and COVID-19 transmission in the USA: A mathematical modelling study. *The Lancet Infectious Diseases*, *20*(11), 1247–1254. https://doi.org/10.1016/S1473-3099(20)30553-3

Bouzouina, L., Kourtit, K., & Nijkamp, P. (2022). Impact of immobility and mobility activities on the spread of COVID-19: Evidence from European countries. *Regional Science Policy & Practice*, *14*(S1), 6–20. https://doi.org/10.1111/rsp3.12565

Bryant, P., & Elofsson, A. (2020). Estimating the impact of mobility patterns on COVID-19 infection rates in 11 European countries. *PeerJ*, *8*, e9879. https://doi.org/10.7717/peerj.9879

Buesa, A., Pérez, J. J., & Santabárbara, D. (2021). Awareness of pandemics and the impact of COVID-19. *Economics Letters*, *204*, 109892. https://doi.org/10.1016/j.econlet.2021.109892

Chang, H.-Y., Tang, W., Hatef, E., Kitchen, C., Weiner, J. P., & Kharrazi, H. (2021). Differential impact of mitigation policies and socioeconomic status on COVID-19 prevalence and social distancing in the United States. *BMC Public Health*, *21*(1), 1140. https://doi.org/10.1186/s12889-021-11149-1

Chen, C. W. S., So, M. K. P., & Liu, F.-C. (2022). Assessing government policies' impact on the COVID-19 pandemic and elderly deaths in East Asia. *Epidemiology & Infection*, *150*, e161. https://doi.org/10.1017/S0950268822001388

Chen, Y., Chen, M., Huang, B., Wu, C., & Shi, W. (2021). Modeling the Spatiotemporal Association Between COVID-19 Transmission and Population Mobility Using Geographically and Temporally Weighted Regression. *GeoHealth*, *5*(5), e2021GH000402. https://doi.org/10.1029/2021GH000402

Cheng, T., & Adepeju, M. (2014). Modifiable Temporal Unit Problem (MTUP) and Its Effect on Space-Time Cluster Detection. *PLOS ONE*, *9*(6), e100465. https://doi.org/10.1371/journal.pone.0100465

Chung, H. W., Apio, C., Goo, T., Heo, G., Han, K., Kim, T., Kim, H., Ko, Y., Lee, D., Lim, J., Lee, S., & Park, T. (2021). Effects of government policies on the spread of COVID-19 worldwide. *Scientific Reports*, *11*(1), Article 1. https://doi.org/10.1038/s41598-021-99368-9

Chung, P.-C., & Chan, T.-C. (2021). Impact of physical distancing policy on reducing transmission of SARS-CoV-2 globally: Perspective from government's response and residents' compliance. *PLOS ONE*, *16*(8), e0255873. https://doi.org/10.1371/journal.pone.0255873

Cinarka, H., Uysal, M. A., Cifter, A., Niksarlioglu, E. Y., & Çarkoğlu, A. (2021). The relationship between Google search interest for pulmonary symptoms and COVID-19 cases using dynamic conditional correlation analysis. *Scientific Reports*, *11*(1), 14387. https://doi.org/10.1038/s41598-021-93836-y

Dainton, C., & Hay, A. (2021). Quantifying the relationship between lockdowns, mobility, and effective reproduction number (Rt) during the COVID-19 pandemic in the Greater Toronto Area. *BMC Public Health*, *21*(1), 1658. https://doi.org/10.1186/s12889-021-11684-x

de la Rosa, P. A., Cowden, R. G., de Filippis, R., Jerotic, S., Nahidi, M., Ori, D., Orsolini, L., Nagendrappa, S., Pinto da Costa, M., Ransing, R., Saeed, F., Shoib, S., Turan, S., Ullah, I., Vadivel, R., & Ramalho, R. (2022). Associations of lockdown stringency and duration with Google searches for mental health terms during the COVID-19 pandemic: A nine-country study. *Journal of Psychiatric Research*, *150*, 237–245. https://doi.org/10.1016/j.jpsychires.2022.03.026

Devaraj, S., & Patel, P. C. (2021). Change in psychological distress in response to changes in reduced mobility during the early 2020 COVID-19 pandemic: Evidence of modest effects from the U.S. *Social Science & Medicine*, *270*, 113615. https://doi.org/10.1016/j.socscimed.2020.113615

Díaz-Castro, L., Cabello-Rangel, H., & Hoffman, K. (2021). The Impact of Health Policies and Sociodemographic Factors on Doubling Time of the COVID-19 Pandemic in Mexico.

*International Journal of Environmental Research and Public Health*, *18*(5), Article 5. https://doi.org/10.3390/ijerph18052354

Effenberger, M., Kronbichler, A., Shin, J. I., Mayer, G., Tilg, H., & Perco, P. (2020). Association of the COVID-19 pandemic with Internet Search Volumes: A Google TrendsTM Analysis. *International Journal of Infectious Diseases*, *95*, 192–197. https://doi.org/10.1016/j.ijid.2020.04.033

Elitzur, M., Kaplan, S., Ivezić, Ž., & Zilberman, D. (2021). The impact of policy timing on the spread of COVID-19. *Infectious Disease Modelling*, *6*, 942–954. https://doi.org/10.1016/j.idm.2021.07.005

Gong, W., Ju, G., Zhu, M., Wang, S., Guo, W., & Chen, Y. (2022). Exploring the Longitudinal Relationship Between Lockdown Policy Stringency and Public Negative Emotions Among 120 Countries During the COVID-19 Pandemic: Mediating Role of Population Mobility. *Frontiers in Psychiatry*, *13*. https://www.frontiersin.org/journals/psychiatry/articles/10.3389/fpsyt.2022.753703

Gordon, D. V., Grafton, R. Q., & Steinshamn, S. I. (2021). Cross-country effects and policy responses to COVID-19 in 2020: The Nordic countries. *Economic Analysis and Policy*, *71*, 198–210. https://doi.org/10.1016/j.eap.2021.04.015

Gottumukkala, R., Katragadda, S., Bhupatiraju, R. T., Kamal, A. Md., Raghavan, V., Chu, H., Kolluru, R., & Ashkar, Z. (2021). Exploring the relationship between mobility and COVID− 19 infection rates for the second peak in the United States using phase-wise association. *BMC Public Health*, *21*(1), 1669. https://doi.org/10.1186/s12889-021-11657-0

Guo, W., Feng, Y., Luo, W., Ren, Y., Tan, J., Jiang, X., & Xue, Q. (2022). The Impacts of COVID-19 and Policies on Spatial and Temporal Distribution Characteristics of Traffic: Two Examples in Beijing. *Sustainability*, *14*(3), Article 3. https://doi.org/10.3390/su14031733

Guo, Y., Yu, H., Zhang, G., & Ma, D. T. (2021). Exploring the impacts of travel-implied policy factors on COVID-19 spread within communities based on multi-source data interpretations. *Health & Place*, *69*, 102538. https://doi.org/10.1016/j.healthplace.2021.102538

Han, Q., Zheng, B., Agostini, M., Bélanger, J. J., Gützkow, B., Kreienkamp, J., Reitsema, A. M., van Breen, J. A., Collaboration, P., & Leander, N. P. (2021). Associations of risk perception of COVID-19 with emotion and mental health during the pandemic. *Journal of Affective Disorders*, *284*, 247–255. https://doi.org/10.1016/j.jad.2021.01.049

He, S., Lee, J., Langworthy, B., Xin, J., James, P., Yang, Y., & Wang, M. (2021). *Delay in the Effect of Restricting Community Mobility on the Spread of COVID-19 in the United States* (SSRN Scholarly Paper 3845372). https://doi.org/10.2139/ssrn.3845372

Hsiang, S., Allen, D., Annan-Phan, S., Bell, K., Bolliger, I., Chong, T., Druckenmiller, H., Huang, L. Y., Hultgren, A., Krasovich, E., Lau, P., Lee, J., Rolf, E., Tseng, J., & Wu, T. (2020). The effect of large-scale anti-contagion policies on the COVID-19 pandemic. *Nature*, *584*(7820), Article 7820. https://doi.org/10.1038/s41586-020-2404-8

Jewell, S., Futoma, J., Hannah, L., Miller, A. C., Foti, N. J., & Fox, E. B. (2021). It's complicated: Characterizing the time-varying relationship between cell phone mobility and COVID-19 spread in the US. *Npj Digital Medicine*, *4*(1), Article 1. https://doi.org/10.1038/s41746-021-00523-3

Jun, S.-P., Yoo, H. S., & Lee, J.-S. (2021). The impact of the pandemic declaration on public awareness and behavior: Focusing on COVID-19 google searches. *Technological Forecasting and Social Change*, *166*, 120592. https://doi.org/10.1016/j.techfore.2021.120592

Jung, S., Endo, A., Akhmetzhanov, A. R., & Nishiura, H. (2021). Predicting the effective reproduction number of COVID-19: Inference using human mobility, temperature, and risk awareness. *International Journal of Infectious Diseases*, *113*, 47–54. https://doi.org/10.1016/j.ijid.2021.10.007

Kallidoni, M., Katrakazas, C., & Yannis, G. (2022). Modelling the relationship between covid-19 restrictive measures and mobility patterns across Europe using time-series analysis. *European Journal of Transport and Infrastructure Research*, *22*(2), Article 2. https://doi.org/10.18757/ejtir.2022.22.2.5728

Kaufman, B. G., Whitaker, R., Mahendraratnam, N., Hurewitz, S., Yi, J., Smith, V. A., & McClellan, M. (2021). State variation in effects of state social distancing policies on COVID-19 cases. *BMC Public Health*, *21*(1), 1239. https://doi.org/10.1186/s12889-021-11236-3

Khan, D., Ahmed, N., Mehmed, B., & Haq, I. ul. (2021). Assessing the Impact of Policy Measures in Reducing the COVID-19 Pandemic: A Case Study of South Asia. *Sustainability*, *13*(20), Article 20. https://doi.org/10.3390/su132011315

Kraemer, M. U. G., Yang, C.-H., Gutierrez, B., Wu, C.-H., Klein, B., Pigott, D. M., Covid, O., Hanage, W. P., Brownstein, J. S., Layan, M., Vespignani, A., Tian, H., Dye, C., Pybus, O. G., & Scarpino, S. V. (2020). *The effect of human mobility and control measures on the COVID-19 epidemic in China*.

Kumar, H., Nataraj, M., & Kundu, S. (2022). COVID-19 and Federalism in India: Capturing the Effects of State and Central Responses on Mobility. *The European Journal of Development Research*, *34*(5), 2463–2492. https://doi.org/10.1057/s41287-021-00463-4

Kuster, A. C., & Overgaard, H. J. (2021). A novel comprehensive metric to assess effectiveness of COVID-19 testing: Inter-country comparison and association with geography, government, and policy response. *PLOS ONE*, *16*(3), e0248176. https://doi.org/10.1371/journal.pone.0248176

Li, Y., Li, M., Rice, M., Zhang, H., Sha, D., Li, M., Su, Y., & Yang, C. (2021). The Impact of Policy Measures on Human Mobility, COVID-19 Cases, and Mortality in the US: A Spatiotemporal Perspective. *International Journal of Environmental Research and Public Health*, *18*(3), Article 3. https://doi.org/10.3390/ijerph18030996

Lin, B., Zou, L., Zhao, B., Huang, X., Cai, H., Yang, M., & Zhou, B. (2024). Sensing the pulse of the pandemic: Unveiling the geographical and demographic disparities of public sentiment toward COVID-19 through social media. *Cartography and Geographic Information Science*, *0*(0), 1–19. https://doi.org/10.1080/15230406.2024.2323489

Lison, A., Persson, J., Banholzer, N., & Feuerriegel, S. (2022). Estimating the effect of mobility on SARS-CoV-2 transmission during the first and second wave of the COVID-19 epidemic,

Switzerland, March to December 2020. *Eurosurveillance*, *27*(10), 2100374. https://doi.org/10.2807/1560-7917.ES.2022.27.10.2100374

Méndez-Lizárraga, C. A., Castañeda-Cediel, Ml., Delgado-Sánchez, G., Ferreira-Guerrero, E. E., Ferreyra-Reyes, L., Canizales-Quintero, S., Mongua-Rodríguez, N., Tellez-Vázquez, N., Jiménez-Corona, M. E., Bradford Vosburg, K., Bello-Chavolla, O. Y., & García-García, L. (2022). Evaluating the impact of mobility in COVID-19 incidence and mortality: A case study from four states of Mexico. *Frontiers in Public Health*, *10*. https://doi.org/10.3389/fpubh.2022.877800

Meng, F., Gong, W., Liang, J., Li, X., Zeng, Y., & Yang, L. (2021). Impact of different control policies for COVID-19 outbreak on the air transportation industry: A comparison between China, the U.S. and Singapore. *PLOS ONE*, *16*(3), e0248361. https://doi.org/10.1371/journal.pone.0248361

Mukherjee, S., & Jain, T. (2022). Impact of COVID-19 on the mobility patterns: An investigation of taxi trips in Chicago. *PLOS ONE*, *17*(5), e0267436. https://doi.org/10.1371/journal.pone.0267436

Nakamoto, D., Nojiri, S., Taguchi, C., Kawakami, Y., Miyazawa, S., Kuroki, M., & Nishizaki, Y. (2022). The impact of declaring the state of emergency on human mobility during COVID-19 pandemic in Japan. *Clinical Epidemiology and Global Health*, *17*, 101149. https://doi.org/10.1016/j.cegh.2022.101149

Nguyen, T. D., Gupta, S., Andersen, M. S., Bento, A. I., Simon, K. I., & Wing, C. (2021). Impacts of state COVID-19 reopening policy on human mobility and mixing behavior. *Southern Economic Journal*, *88*(2), 458–486. https://doi.org/10.1002/soej.12538

Paternina-Caicedo, A., Alvis-Guzmán, N., Dueñas, C., Narvaez, J., Smith, A. D., & De la Hoz-Restrepo, F. (2022). Impact of mobility restrictions on the dynamics of transmission of COVID-19 in Colombian cities. *International Health*, *14*(3), 332–335. https://doi.org/10.1093/inthealth/ihab064

Peng, X., Huang, J., Liang, K., & Chi, X. (2022). The Association of Social Emotions, Perceived Efficiency, Transparency of the Government, Concerns about COVID-19, and Confidence in Fighting the Pandemic under the Week-Long Lockdown in Shenzhen, China. *International*

*Journal of Environmental Research and Public Health*, *19*(18), Article 18. https://doi.org/10.3390/ijerph191811173

Pfeiffer, B., Brusilovskiy, E., Hallock, T., Salzer, M., Davidson, A. P., Slugg, L., & Feeley, C. (2022). Impact of COVID-19 on Community Participation and Mobility in Young Adults with Autism Spectrum Disorders. *Journal of Autism and Developmental Disorders*, *52*(4), 1553–1567. https://doi.org/10.1007/s10803-021-05054-0

Poppe, A., & Maskileyson, D. (2022). The effect of non-pharmaceutical policy interventions on COVID-19 transmission across three cities in Colombia. *Frontiers in Public Health*, *10*. https://www.frontiersin.org/articles/10.3389/fpubh.2022.937644

Rahman, M. M., & Thill, J.-C. (2022). Associations between COVID-19 Pandemic, Lockdown Measures and Human Mobility: Longitudinal Evidence from 86 Countries. *International Journal of Environmental Research and Public Health*, *19*(12), Article 12. https://doi.org/10.3390/ijerph19127317

Razzaq, A., Abbas, T., Hashim, S., Qadri, S., Mumtaz, I., Saher, N., Ul-Rehman, M., Shahzad, F., & Nawaz, S. A. (2022). Extraction of Psychological Effects of COVID-19 Pandemic through Topic-Level Sentiment Dynamics. *Complexity*, *2022*, e9914224. https://doi.org/10.1155/2022/9914224

Sibley, C. G., Greaves, L. M., Satherley, N., Wilson, M. S., Overall, N. C., Lee, C. H. J., Milojev, P., Bulbulia, J., Osborne, D., Milfont, T. L., Houkamau, C. A., Duck, I. M., Vickers-Jones, R., & Barlow, F. K. (2020). Effects of the COVID-19 pandemic and nationwide lockdown on trust, attitudes toward government, and well-being. *American Psychologist*, *75*, 618–630. https://doi.org/10.1037/amp0000662

Sözen, M. E., Sarıyer, G., & Ataman, M. G. (2022). Big data analytics and COVID-19: Investigating the relationship between government policies and cases in Poland, Turkey and South Korea. *Health Policy and Planning*, *37*(1), 100–111. https://doi.org/10.1093/heapol/czab096

Sukhwal, P. C., & Kankanhalli, A. (2022). Determining containment policy impacts on public sentiment during the pandemic using social media data. *Proceedings of the National Academy of Sciences*, *119*(19), e2117292119. https://doi.org/10.1073/pnas.2117292119

Tao, S., Bragazzi, N. L., Wu, J., Mellado, B., & Kong, J. D. (2022). Harnessing Artificial Intelligence to assess the impact of nonpharmaceutical interventions on the second wave of the Coronavirus Disease 2019 pandemic across the world. *Scientific Reports*, *12*(1), Article 1. https://doi.org/10.1038/s41598-021-04731-5

Tokey, A. I. (2021). Spatial association of mobility and COVID-19 infection rate in the USA: A county-level study using mobile phone location data. *Journal of Transport & Health*, *22*, 101135. https://doi.org/10.1016/j.jth.2021.101135

Tu, B., Wei, L., Jia, Y., & Qian, J. (2021). Using Baidu search values to monitor and predict the confirmed cases of COVID-19 in China: – Evidence from Baidu index. *BMC Infectious Diseases*, *21*(1), 98. https://doi.org/10.1186/s12879-020-05740-x

Vega-Villalobos, A., Almanza-Ortega, N. N., Torres-Poveda, K., Pérez-Ortega, J., & Barahona, I. (2022). Correlation between mobility in mass transport and mortality due to COVID-19: A comparison of Mexico City, New York, and Madrid from a data science perspective. *PLOS ONE*, *17*(3), e0264713. https://doi.org/10.1371/journal.pone.0264713

Vinceti, M., Balboni, E., Rothman, K. J., Teggi, S., Bellino, S., Pezzotti, P., Ferrari, F., Orsini, N., & Filippini, T. (2022). Substantial impact of mobility restrictions on reducing COVID-19 incidence in Italy in 2020. *Journal of Travel Medicine*, *29*(6), taac081. https://doi.org/10.1093/jtm/taac081

Wallin Aagesen, H., Järv, O., & Gerber, P. (2022). The effect of COVID-19 on cross-border mobilities of people and functional border regions: The Nordic case study from Twitter data. *Geografiska Annaler: Series B, Human Geography*, *0*(0), 1–23. https://doi.org/10.1080/04353684.2022.2101135

Wang, S., Liu, Y., & Hu, T. (2020). Examining the Change of Human Mobility Adherent to Social Restriction Policies and Its Effect on COVID-19 Cases in Australia. *International Journal of*

*Environmental Research and Public Health*, *17*(21), Article 21. https://doi.org/10.3390/ijerph17217930

Wang, S., Tong, Y., Fan, Y., Liu, H., Wu, J., Wang, Z., & Fang, C. (2021). Observing the silent world under COVID-19 with a comprehensive impact analysis based on human mobility. *Scientific Reports*, *11*(1), Article 1. https://doi.org/10.1038/s41598-021-94060-4

Wellenius, G. A., Vispute, S., Espinosa, V., Fabrikant, A., Tsai, T. C., Hennessy, J., Dai, A., Williams, B., Gadepalli, K., Boulanger, A., Pearce, A., Kamath, C., Schlosberg, A., Bendebury, C., Mandayam, C., Stanton, C., Bavadekar, S., Pluntke, C., Desfontaines, D., … Gabrilovich, E. (2021). Impacts of social distancing policies on mobility and COVID-19 case growth in the US. *Nature Communications*, *12*(1), Article 1. https://doi.org/10.1038/s41467-021-23404-5

Widyasari, V., Lee, C. B., Lin, K.-H., Husnayain, A., Su, E. C.-Y., & Wang, J.-Y. (2022). Effects of the Government Response and Community Mobility on the COVID-19 Pandemic in Southeast Asia. *Healthcare*, *10*(10), Article 10. https://doi.org/10.3390/healthcare10102003

Wong, D. W. S. (2004). The Modifiable Areal Unit Problem (MAUP). In D. G. Janelle, B. Warf, & K. Hansen (Eds.), *WorldMinds: Geographical Perspectives on 100 Problems: Commemorating the 100th Anniversary of the Association of American Geographers 1904–2004* (pp. 571–575). Springer Netherlands. https://doi.org/10.1007/978-1-4020-2352-1_93

Wu, L., & Shimizu, T. (2022). Analysis of the impact of non-compulsory measures on human mobility in Japan during the COVID-19 pandemic. *Cities*, *127*, 103751. https://doi.org/10.1016/j.cities.2022.103751

Xiong, C., Hu, S., Yang, M., Luo, W., & Zhang, L. (2020). Mobile device data reveal the dynamics in a positive relationship between human mobility and COVID-19 infections. *Proceedings of the National Academy of Sciences*, *117*(44), 27087–27089. https://doi.org/10.1073/pnas.2010836117

Yang, B., Huang, A. T., Garcia-Carreras, B., Hart, W. E., Staid, A., Hitchings, M. D. T., Lee, E. C., Howe, C. J., Grantz, K. H., Wesolowksi, A., Lemaitre, J. C., Rattigan, S., Moreno, C., Borgert, B. A., Dale, C., Quigley, N., Cummings, A., McLorg, A., LoMonaco, K., … Cummings, D. A. T.

(2021). Effect of specific non-pharmaceutical intervention policies on SARS-CoV-2 transmission in the counties of the United States. *Nature Communications*, *12*(1), Article 1. https://doi.org/10.1038/s41467-021-23865-8

Zeng, C., Zhang, J., Li, Z., Sun, X., Olatosi, B., Weissman, S., & Li, X. (2021). Spatial-Temporal Relationship Between Population Mobility and COVID-19 Outbreaks in South Carolina: Time Series Forecasting Analysis. *Journal of Medical Internet Research*, *23*(4), e27045. https://doi.org/10.2196/27045

Zhang, J., Zhang, R., Ding, H., Li, S., Liu, R., Ma, S., Zhai, B., Kashima, S., & Hayashi, Y. (2021). Effects of transport-related COVID-19 policy measures: A case study of six developed countries. *Transport Policy, 110,* 37–57. https://doi.org/10.1016/j.tranpol.2021.05.013

## Acknowledgements

This study is based on work supported by the Data Resource Develop Program Award from the Texas A&M Institute of Data Science (TAMIDS). Any opinions, findings, and conclusions or recommendations expressed in this material are those of the authors and do not necessarily reflect the views of the funding agencies.

## Data Availability Statement

The data used in this research were derived from the Web of Science database.

## Disclosure statement

No potential conflict of interest was reported by the authors.